\documentclass[12pt,preprint]{aastex}

\shorttitle{Red Galaxy Clustering in the NDWFS}
\shortauthors{Brown {\it et al.}}

\begin{document}

\title{Red Galaxy Clustering in the NOAO Deep Wide-Field Survey}

\author{Michael J. I. Brown, Arjun Dey, Buell T. Jannuzi, Tod R. Lauer,}
\affil{National Optical Astronomy Observatory, Tucson, AZ 85726-6732}
\email{mbrown@noao.edu, dey@noao.edu, jannuzi@noao.edu, lauer@noao.edu}

\author{Glenn P. Tiede, and Valerie J. Mikles}
\affil{Department of Astronomy, University of Florida, Gainesville, FL 32611-2055}
\email{tiede@astro.ufl.edu, mikles@astro.ufl.edu}

\begin{abstract}
We have measured the clustering of $0.30<z<0.90$ red galaxies 
and constrained models of the evolution of large-scale structure 
using the initial $1.2\Box^\circ$ data release of the NOAO Deep 
Wide-Field Survey (NDWFS). The area and $B_WRI$ passbands of the 
NDWFS allow samples of $\gtrsim 10^3$ galaxies to be selected as a 
function of spectral type, absolute magnitude, and photometric redshift. 
Spectral synthesis models can be used to predict the colors and 
luminosities of a galaxy population as a function of redshift. 
We have used PEGASE2 models, with exponentially declining star formation rates, 
to estimate the observed colors and luminosity evolution of galaxies and 
to connect, as an evolutionary sequence, related populations of 
galaxies at different redshifts. A red galaxy sample, with present-day rest-frame 
Vega colors of $B_W-R>1.44$, was chosen to allow comparisons with the 
2dF Galaxy Redshift Survey and Sloan Digital Sky Survey. 
We find the spatial clustering 
of red galaxies to be a strong function of luminosity, with $r_0$ increasing 
from $4.4\pm 0.4 h^{-1} {\rm Mpc}$ at $M_R-5{\rm log}h\approx -20.0$ to 
$11.2\pm 1.0 h^{-1} {\rm Mpc}$ at $M_R-5{\rm log}h\approx -22.0$. 
Clustering evolution measurements using samples where the 
rest-frame selection criteria vary with redshift, 
including all deep single-band magnitude limited samples, 
are biased due to the correlation of clustering with rest-frame 
color and luminosity. The clustering of $-21.5<M_R-5{\rm log}h<-20.5$,
$B_W-R>1.44$ galaxies exhibits no significant evolution over 
the redshift range observed with $r_0= 6.3\pm 0.5 h^{-1} {\rm Mpc}$ in
comoving coordinates. This is consistent with recent $\Lambda$CDM 
models where the bias of $L^*$ galaxies undergoes rapid evolution 
and $r_0$ evolves very slowly at $z<2$. 
\end{abstract}


\section{Introduction}
\label{sec:intro}

The strength and evolution of the spatial clustering of galaxies 
is a function of cosmology, galaxy mass, and galaxy formation
scenarios in Cold Dark Matter models \citep[e.g.,][]{pea97,kau99,ben01,som01}. 
The predicted linear and quasi-linear growth of density perturbations results
in the spatial clustering of dark matter undergoing rapid evolution 
at $z<2$ \citep[e.g.,][]{pea97}. However, CDM semi-analytic
models and simulations predict the distribution of $M_R<-20$ 
galaxies to be highly biased with respect to the underlying dark 
matter distribution \citep[e.g.,][]{col89,kau99,ben01,som01}. 
CDM models for an $\Omega_m=0.3$, $\Lambda=0.7$ cosmology ($\Lambda$CDM)
predict little evolution of the early-type galaxy spatial 
correlation function (measured in comoving coordinates) at $z<2$.

At low redshift, strong constraints on galaxy clustering as a function of 
spectral type and absolute magnitude are provided by the 2dF Galaxy Redshift Survey 
(2dFGRS) and Sloan Digital Sky Survey (SDSS) wide-field 
spectroscopic surveys \citep{nor02,zeh02}. At $z>0.3$,
the comoving volume and galaxy counts of spectroscopic surveys are orders of 
magnitude less than the 2dFGRS and SDSS and the resulting estimates of the 
spatial correlation function suffer from small number statistics
and biases from unrepresentative sample volumes 
\citep{hog00,gia01}. Only wide-field surveys, with large samples ($>10^3$) of galaxies
and volumes of $\gtrsim10^6 h^{-3} {\rm Mpc^3}$, are able to provide robust estimates
of $r_0$ (the spatial scale where the correlation function equals 1) with 
errors of $<10\%$.

It is possible to obtain constraints on the clustering 
of high-redshift galaxies using the two-point angular correlation
function and deep wide-field imaging. While survey volumes at high
redshift have increased with the advent of large-format CCDs,
significant issues remain in the interpretation 
of angular correlation functions derived from these samples.
Estimates of the spatial correlation function derived from imaging  
rely on models of the redshift distribution which differ 
significantly from each other. In addition, the distribution of 
galaxy types and luminosities can be a strong function of apparent 
magnitude and redshift. As the clustering of galaxies is a function of 
spectral type and luminosity \citep[e.g.,][]{dav76,lov95,nor02,zeh02}, 
the apparent evolution of the correlation function can be dominated by 
selection effects rather than evolution of large-scale structure \citep{efs91}.

The advent of reliable photometric redshifts allows improved 
constraints on the redshift distribution of faint galaxies in imaging 
surveys \citep[e.g., Brunner, Szalay \& Connolly 2000;][]{bro01,tep01,fir02}. 
However, as discussed in \S\ref{sec:cor}, the resulting model redshift 
distributions and spatial correlation functions strongly depend on the 
uncertainties of photometric redshifts. If rest-frame color criteria are 
also applied, samples of galaxies containing a comparable range of spectral types can be 
selected over a broad range of redshifts \citep[Brown et al. 2001,][]{fir02,wil03}, 
thus reducing the selection effect which dominates clustering evolution 
estimates derived from single-band imaging. Absolute magnitude selection 
criteria can also be used, though this requires accurate photometric redshifts.

To measure the evolution of galaxy clustering, we have used optical
images from the NDWFS \citep[][Jannuzi et al., in preparation]{jan99} 
to select comparable samples of red galaxies at multiple 
epochs. We have used photometric redshifts and the 
PEGASE2 galaxy spectral evolution models \citep{fio97} with exponentially
declining star formation rates to estimate the redshifts, luminosities, 
and rest-frame colors of all 
galaxies detected in the $B_W$, $R$, and $I$-bands. In addition, we used the 
best-fit  PEGASE2 models to estimate the spectral evolution and 
the present-day rest-frame colors and luminosities.
We selected a red galaxy sample, with present-day rest-frame colors 
of $B_W-R>1.44$, to allow direct comparison of the NDWFS clustering 
measurements with the low redshift early-type galaxy clustering 
measurements from the 2dFGRS. 
We also selected a $B_W-R>1.77$ subsample
to allow a measurement of galaxy clustering as a function rest-frame color.

The outline of the paper is as follows. We provide a brief description of the NDWFS 
imaging data and catalogs in \S\ref{sec:ndwfs}. We describe the estimation of photometric 
redshifts and the galaxy spectral evolution models in \S\ref{sec:photoz}. In \S\ref{sec:red}, we 
discuss the motivation for studying red galaxies and the selection of these objects. 
We discuss the measurement of the angular and spatial two-point correlation functions in 
\S\ref{sec:cor}.  In \S\ref{sec:clust}, we present the clustering of $0.30<z<0.90$ 
red galaxies as a function of absolute magnitude and redshift. We discuss the implications
of our results in \S\ref{sec:dis} and summarize the paper in \S\ref{sec:sum}.

\section{The NOAO Deep Wide-Field Survey}
\label{sec:ndwfs}

The NDWFS is a multiband ($B_W,R,I,J,H,K$) survey of two 
$\approx 9\Box^\circ$ high Galactic latitude fields with the CTIO 
$4{\rm m}$, KPNO $4 {\rm m}$, and KPNO $2.1 {\rm m}$ telescopes \citep{jan99}. 
A thorough description of the observing strategy and data reduction 
will be provided by Jannuzi et al. (in preparation). This paper utilizes 
the first $1.2\Box^\circ$ $B_WRI$ public data release of four adjacent 
KPNO $4 {\rm m}$ MOSAIC subfields in the Bo\"{o}tes field.
The coordinates, depth, and image quality of the four subfields are provided in 
Table~\ref{table:fields}.
This dataset, along with explanatory material,  is available from the 
NOAO Science Archive on the World Wide Web\footnote{http://www.archive.noao.edu/ndwfs/}. 

We generated object catalogs using SExtractor $2.2.2$ \citep{ber96} 
run in single-image mode with the minimum detection area, convolution filter, 
and signal above sky threshold optimized to provide the deepest catalogs possible 
as a function of the seeing. Detections in the different bands were then 
matched using criteria based on the distance between the image centroids. 
At faint magnitudes ($R>22$), images with centroids within $1^{\prime\prime}$ 
of each other in different bands were matched. At bright magnitudes, 
images were matched if the probability of the centroids being within 
a given distance of each other by random chance was $<0.005$. 
This probability was determined using the number of objects
as a function of magnitude per unit area measured from the NDWFS.
Where multiple matches were found, the closest centroids were matched. 
In practice, such criteria work well for both faint objects and bright 
objects including saturated stars and well resolved galaxies.

We have used SExtractor MAG\_AUTO magnitudes \citep{ber96}, which 
are similar to Kron total magnitudes \citep{kro80}, due to their 
low uncertainties and small systematic errors at faint magnitudes. The 
$1\sigma$ uncertainty of the MAG\_AUTO photometry decreases 
from $\simeq 0.3$ magnitudes at the $50\%$ completeness limit to  
$\simeq 0.1$ magnitudes at 2 magnitudes brighter than the completeness 
limit. For unsaturated objects, the MAG\_AUTO photometry has systematic errors
which are significantly less than the $1\sigma$ uncertainties.
The measured correlation functions do not strongly depend on the 
technique used to measure object fluxes. Correlation functions 
using galaxies selected with MAG\_AUTO photometry
and $5^{\prime\prime}$ aperture photometry differed by $\lesssim 1 \sigma$.
Throughout the remainder of the paper, all the results and conclusions 
are derived from NDWFS samples using  MAG\_AUTO photometry.

We determined the completeness as a function of magnitude by adding artificial 
stellar objects to copies of the data and recovering them with SExtractor. 
The $50\%$ completeness limits vary between the 4 subfields in the ranges of 
$26.2<B_W<26.8$, $24.8<R<25.3$, and $23.6<I<25.6$. We used SExtractor's 
star-galaxy classifier to remove objects from the galaxy catalog which had a 
stellarity of $>0.7$ in 2 or more bands brighter than $B_W<23.8$, 
$R<22.8$, and $I<21.4$. Tests with artificial stars indicate 
$\lesssim 5 \%$ of stars are being misclassified as galaxies at the 
classification limit in all four subfields.

Regions surrounding saturated stars were removed from the catalog to
exclude clustered spurious objects detected in the wings of the point
spread function of bright stars. Unclustered contamination reduces the amplitude of 
the correlation function by $(1-f)^2$ where $f$ is the fractional contamination of 
the sample. We assume that $1\%$ of bright stars are misclassified 
as galaxies, a rate which is consistent with tests with artificial 
stars which are a magnitude brighter than the classification limits. 
While this is a crude assumption, misclassified bright
stars would only significantly alter the correlation function 
results if they were misclassified at a rate of more than $10\%$.

The number of objects classified as stars rapidly decreases at 
magnitudes fainter than $R\approx 22$. To estimate the number 
counts of faint red stars, we assumed stars fainter than the classification limits 
have the same color distribution as stars within a magnitude of the 
classification limits. We then modeled the star counts 
with a power law given by
\begin{equation}
\frac{dN}{dm} \propto 10^{\alpha m}.
\end{equation}
This relationship was normalized so the counts agreed with the measured
star counts at the classification limits. For this work, we assumed 
$\alpha=0.07$. While this is only an approximation, the increase in the counts 
is comparable to the observed star counts in the Medium Deep Survey \citep{san96}. 
It is also comparable to the power-law index for the counts of $19<R<21$
red stars in the NDWFS. Tests with $\alpha$ between $0.0$ and $0.15$ did not alter the amplitude of 
the angular correlation function significantly. In addition, the amplitude of the 
correlation function (after correcting for contamination) does not significantly
change if the star-galaxy classification magnitude limits are decreased. For all the correlation 
function bins, the estimated rate of stellar contamination was less than $12\%$.

\section{Photometric redshifts}
\label{sec:photoz}

Photometric redshifts were determined for all objects with $B_W$, $R$, and 
$I$-band detections. Model spectra as a function of redshift were 
generated using PEGASE2 spectral synthesis code \citep{fio97}. 
Models with $z=0$ Solar metallicity, Miller-Scalo initial mass functions, 
$z=0$ ages of $12 ~{\rm Gyr}$ (formation $z \approx 4.1$), and exponentially 
decreasing star formation rates with $e$-folding times between $0.6~{\rm Gyr}$ 
and $15~{\rm Gyr}$ ($\tau$-models) for an $\Omega=0.3$ $\Lambda=0.7$ 
$H_0=70~{\rm km~s}^{-1} ~{\rm Mpc}^{-1}$ cosmology\footnote{Throughout this 
paper $h \equiv \frac{H_0}{100 {\rm km~s}^{-1} ~{\rm Mpc}^{-1}}$, 
$\Omega_m=0.3$, and $\Lambda=0.7$.} were used to estimate galaxy colors, $k$-corrections 
and spectral evolution corrections. At $z>3.9$, $\tau$-models
with ages of $100~{\rm Myr}$ were used to model the spectra of high redshift
galaxies. The model spectra were multiplied by intrinsic dust extinction with 
$E(B-V)=0.04$ and $R_V=3.1$, comparable to estimates for $0<z<1$ early-type 
galaxies \citep{fal99,kau03}. We multiplied the $B_W$, $R$, and $I$ filter 
transmission curves with the MOSAIC CCD quantum efficiency as a function of wavelength, 
the mirror reflectivity and a measurement of the KPNO atmospheric extinction to 
improve the accuracy of the model galaxy colors. Galaxy photometry was corrected for 
Galactic dust extinction using the dust maps of \cite{sch98}, though it should be noted 
that the maximum estimate of $E(B-V)$ was only $0.014$ in the four subfields. We used the 
Vega spectral energy distribution of \cite{hay85} to zeropoint the model galaxy colors. 
Model galaxy colors zeropointed with the \cite{hay85} spectrum or the \cite{cas94} model 
of Vega differ by $B_W-R\simeq 0.03$ from model galaxy colors zeropointed with the 
frequently used \cite{kur79} model of Vega. Interpolation between the $\tau$-model was 
used to fill the color-space occupied by galaxies observed in the 
NDWFS. The uncertainties of the photometric redshifts would be underestimated if 
the color-space was not filled. This is particularly true if color-redshift 
degeneracies present in the data are not reproduced by the models.

\begin{figure}
\plottwo{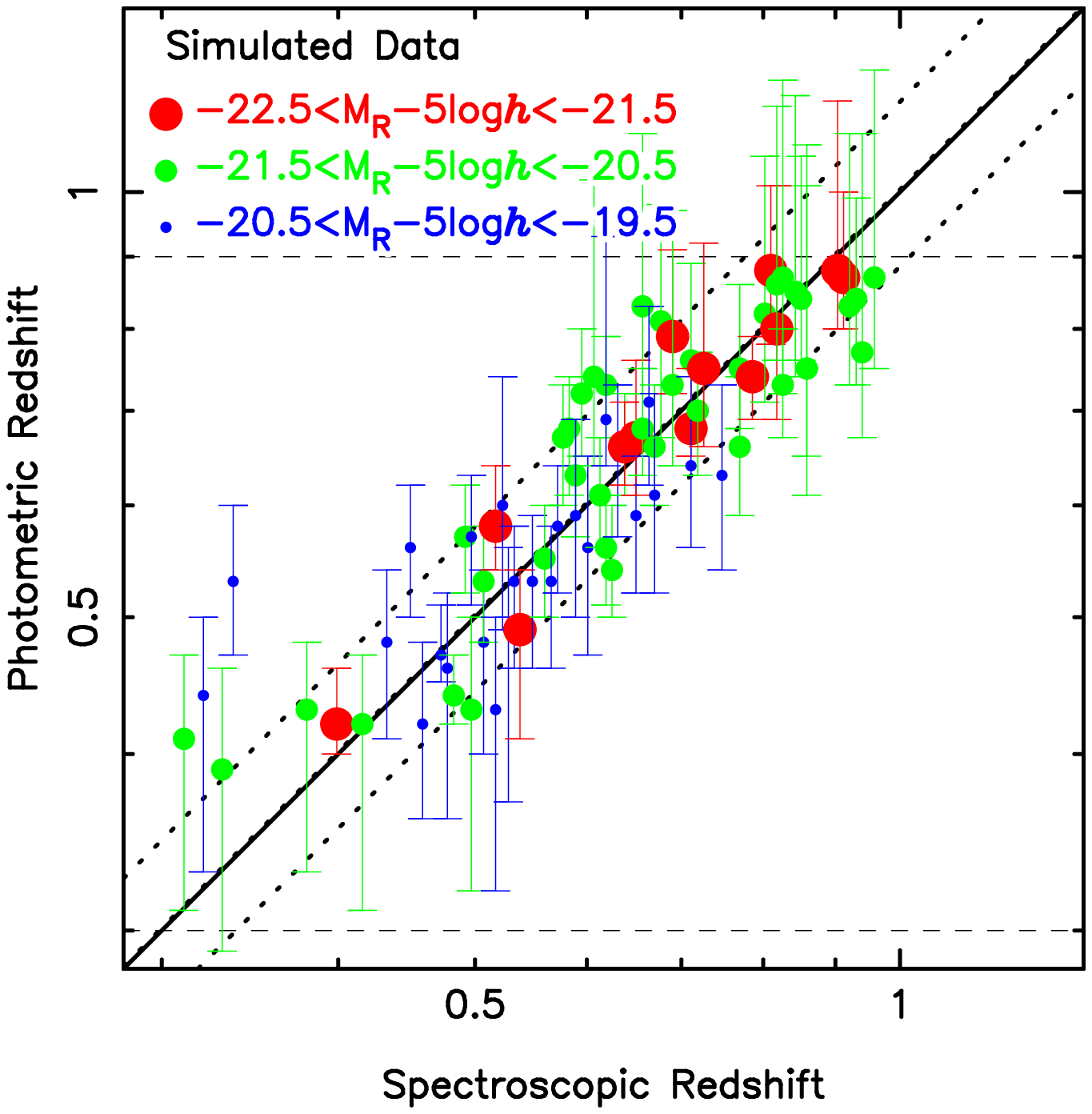}{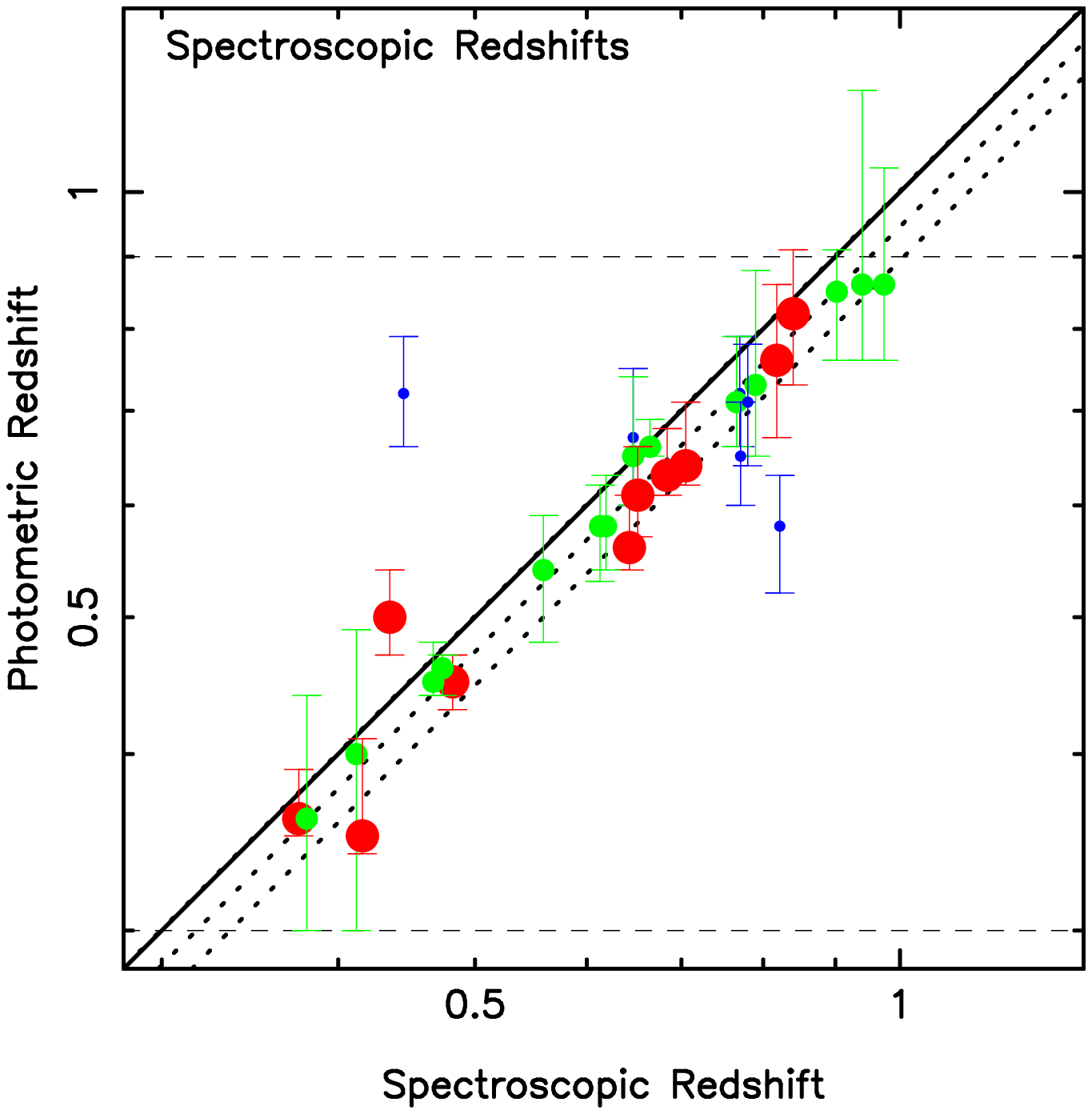}
\caption{A comparison of the photometric and spectroscopic redshifts of $-22.5<M_R-5{\rm log}h<-19.5$ 
red galaxies ($\tau<4.5~{\rm Gyr}$) with photometric redshifts in the range $0.30<z<0.90$. 
Dashed horizontal lines mark the upper and lower limits of the photometric redshift range.
Dotted diagonal lines show the median offset and $\pm 1 \sigma$ uncertainty of the photometric redshifts.
Simulated galaxies, generated using the $\tau$-models with photometric noise added, are 
shown in the lefthand panel. On the right are real galaxies with spectroscopic 
redshifts from the NED database and unpublished spectra.
The real galaxies show a $5\%$ systematic offset between the photometric and spectroscopic 
redshifts. With the inclusion of 0.05 magnitudes of intrinsic scatter, the real data have
$1\sigma$ uncertainty estimates which are consistent with the measured scatter
between the photometric and spectroscopic redshifts.}
\label{fig:photoz}
\end{figure}

Photometric redshifts were estimated by finding the minimum value of $\chi^2$
as a function of redshift, spectral type ($\tau$), and luminosity.  To reduce the CPU time 
required to evaluate the photometric redshifts, the interpolated models were only 
evaluated when they differed from neighboring models sufficiently to significantly alter
the photometric redshifts. As the model spectra do not account for the observed width
of the galaxy locus, we increased the photometric uncertainties for the galaxies by 
$0.05$ magnitudes (added in quadrature). This results in the photometric redshift code
producing $1\sigma$ errors which are consistent with the observed scatter between the 
photometric and spectroscopic redshifts in Figure~\ref{fig:photoz}. 

To further improve the accuracy of the photometric redshifts and their
uncertainties, the estimated redshift distribution of galaxies as a function of 
spectral type and apparent magnitude was introduced as a prior. 
This approach, rather than a Bayesian prior \citep{kod99,ben00}, 
was undertaken, as the number of objects as a function of $\tau$ is poorly known 
at present. The 2dFGRS $z=0$ luminosity functions for different spectral types 
\citep{mad02}, with luminosity evolution given by the $\tau$-models, were used to 
produce estimates of the redshift distributions. At $M_R-5{\rm log}h<-23$, 
where the luminosity function of red galaxies is poorly fitted by a Schechter function 
\citep{mad02}, the luminosity function was approximated by a power-law.
To obtain the correspondence between the 2dFGRS principal component $\eta$ 
parameter and the $\tau$ parameter, we fitted PEGASE2 models to the
2dFGRS principal component spectra between 3900\AA ~and 4100\AA. 
We did not fit for the entire 2dFGRS spectrum as there may be errors in the 
2dFGRS continuum calibration \citep{mad02}. 
The prior has little effect on the best-fit estimates of the photometric redshifts
but does significantly alter the photometric redshift uncertainties. 

To confirm the reliability of the photometric redshifts, spectral types, and 
absolute magnitudes, simulated data were generated using the 
$\tau$ spectral evolution models and the 2dFGRS luminosity functions. 
The simulated data consisted of $\tau \leq 15~{\rm Gyr}$ galaxies with 
the redshift range $0<z\leq5$ and luminosity range $0.01<L^*\leq 100$. 
The simulated object photometry was scattered using the estimated 
uncertainties (including the 0.05 component discussed earlier) as a function of
apparent magnitude, thus mimicking what would be present in the real catalogs.
In addition to the simulated data, the accuracy of the photometric redshifts
was confirmed with  spectroscopic redshifts and $B_WRI$ photometry for 
selected objects in the NDWFS Bo\"{o}tes, NDWFS Cetus, and Lockman Hole fields.

\begin{figure}
\plottwo{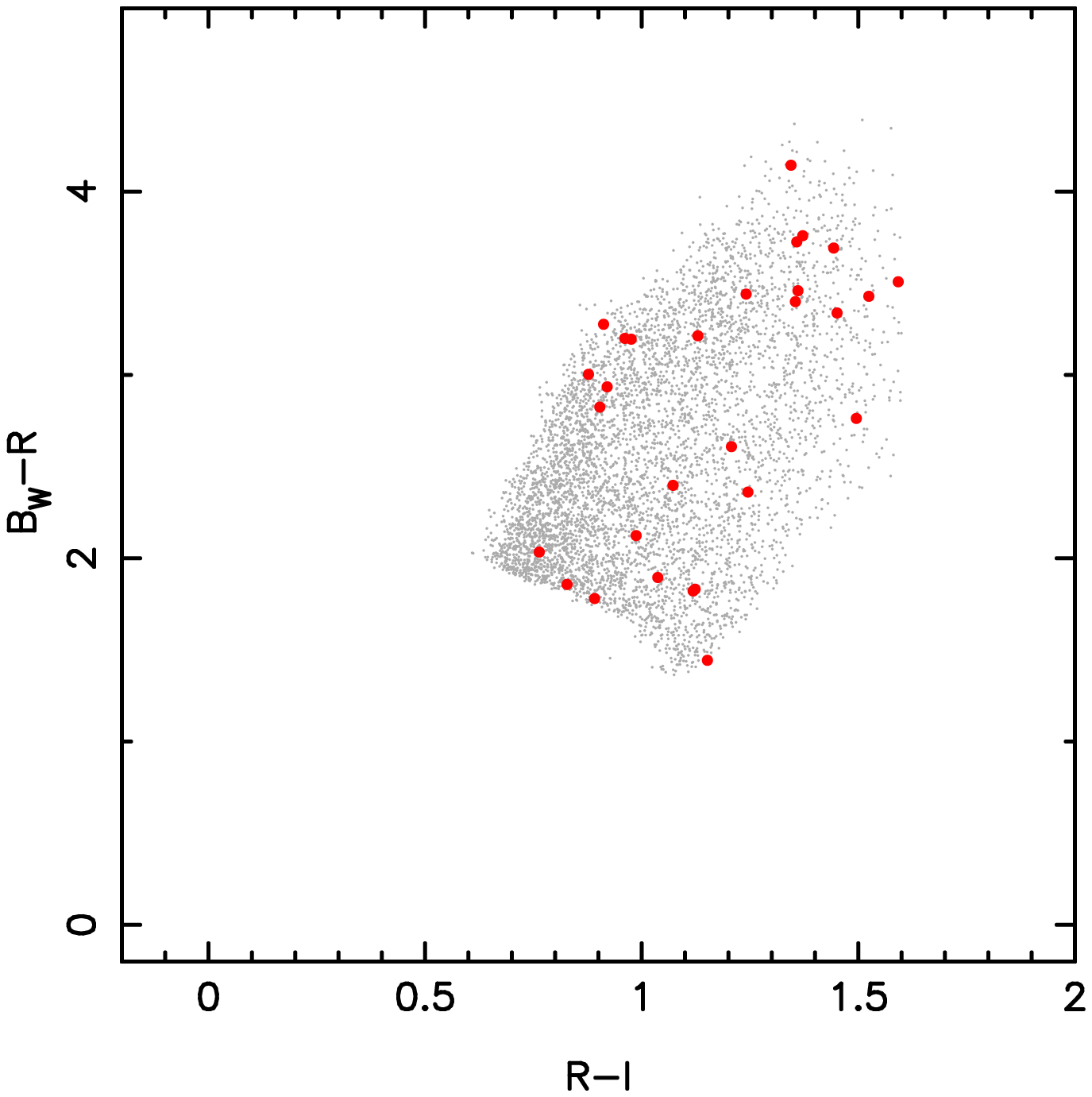}{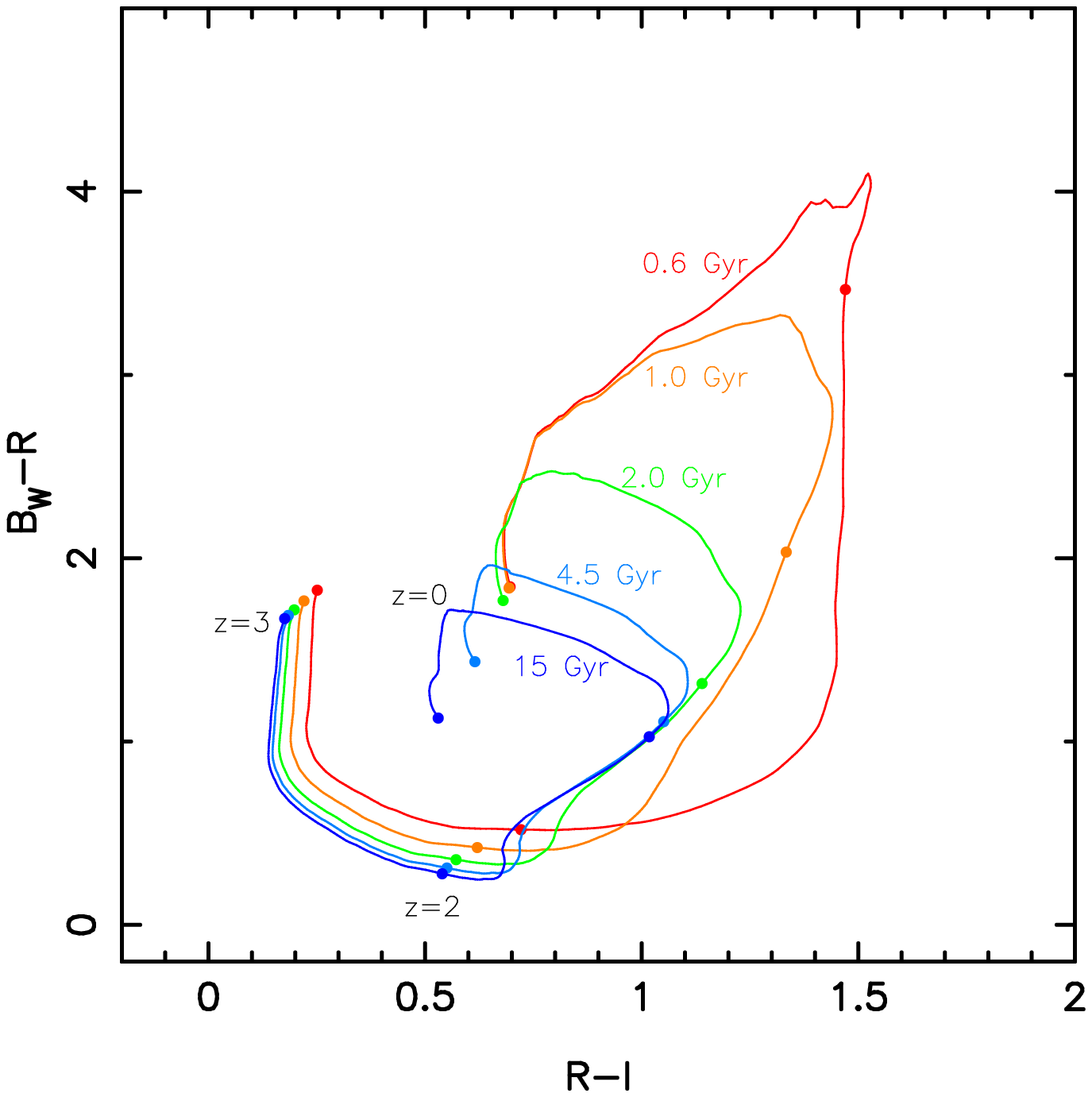}
\caption{The Vega colors of the $\tau<4.5~{\rm Gyr}$, $0.30<z<0.90$ NDWFS Bo\"{o}tes 
galaxy sample (left panel) and the $\tau$-models (right panel). Bold points in the 
left hand panel show the colors of galaxies with spectroscopic redshifts 
(including Cetus and Lockman Hole galaxies) which were used to test the accuracy 
of the photometric redshifts. Bold points on the model tracks mark the 
predicted colors of $z=0$, 1, 2 and 3 galaxies for each of the models.
For clarity, only five of the $\tau$-model tracks are plotted. 
The $\tau$-models occupy the same region of color space as the NDWFS red galaxies.
Single burst models (not shown) have different colors as a function of redshift.
A more detailed description of the colors of NDWFS galaxies and the accuracy of galaxy stellar 
evolution models will be provided by Dey et al. (in preparation).}
\label{fig:col}
\end{figure}

A comparison of the photometric and spectroscopic redshifts for 
$-22.5<M_R-5{\rm log}h<-19.5$ red galaxies is shown in right-hand panel of Figure~\ref{fig:photoz}.
The observed colors of the galaxies with spectroscopic redshifts are shown in Figure~\ref{fig:col}.
We discuss the selection criteria for the red galaxies in \S\ref{sec:red}.
The photometric redshifts have a $5\%$ systematic error and $\pm5\%$ $1\sigma$
uncertainty. Only two of the 29 galaxies have photometric redshifts with errors of
$>30\%$. One of the outliers has strong [OII] emission and Balmer absorption.
The other is a blended object consisting of an emission line galaxy and an M-star.

Increasing or decreasing the internal dust extinction increases the
offset between the photometric and spectroscopic redshifts and increases the 
random scatter. As varying the dust extinction does not improve the accuracy
of the photometric redshifts, we use the constant value of $E(B-V)=0.04$.
Photometric redshifts using the GISSEL01 \citep{bru93,liu00} $\tau$-model 
spectral energy distributions produce a slightly larger residual offset
and more random scatter, so we use the PEGASE2 models throughout the paper. 
However, there are no large differences between 
correlation functions determined using galaxies with PEGASE2 and GISSEL01 
photometric redshifts. Throughout the remainder of the paper, photometric 
redshifts and absolute magnitudes have been corrected for the $5\%$ 
systematic error shown in Figure~\ref{fig:photoz}.

\section{The red galaxy sample}
\label{sec:red}

If the PEGASE2 $\tau$-models are a good approximation for the spectral 
evolution of $z<0.90$ galaxies, the spectral model fits by the photometric 
redshift code can be used to measure the spectral types and luminosity 
evolution of galaxies. This assumption is consistent with the $B_WRI$ 
colors of the galaxy locus (Dey et al., in preparation) and comparisons of 
photometric and spectroscopic redshifts discussed in \S\ref{sec:photoz}. 
We were therefore able to select comparable populations 
of galaxies at multiple epochs.

If a galaxy sample is going to be used to measure the evolution of 
clustering, an evolutionary sequence of related galaxy populations 
must be selected at different redshifts. As the clustering of 
galaxies is known to be a function of absolute magnitude at low redshift 
\citep[e.g.,][]{nor02,zeh02}, accurate photometric redshifts are 
required so objects with similar luminosities can be selected at 
multiple epochs, enabling unbiased studies of the evolution of clustering. 
Galaxy types with low rates of spectral evolution are advantageous as 
selection criteria relying on rapidly evolving models will be very 
sensitive to errors in the model spectral energy distributions.
Red galaxies have accurate photometric redshifts and low rates of spectral evolution.
At $z<1$, the 4000\AA~ break is moving through the optical so accurate 
photometric redshifts can be obtained with a limited number of optical passbands. 
The PEGASE2 $\tau\sim 1~{\rm Gyr}$ models and observations \citep{jor99,sch99,im02} 
indicate $z=0.9$ red galaxies are only $\approx 0.9$ magnitudes 
brighter in rest-frame $R$-band than $z=0$ red galaxies.

For this work, we selected a sample of red galaxies to be
comparable to the early-type sample of the 2dFGRS \citep{nor02}.
Galaxies fitted with $\tau<4.5 {\rm ~Gyr}$ were chosen so their
$z=0$ spectra matched the 2dFGRS principal component selection 
criterion. The rest-frame color of the $12~{\rm Gyr}$ old 
$\tau=4.5 {\rm ~Gyr}$  model with $E(B-V)=0.04$ intrinsic dust 
extinction is $B_W-R=1.44$. This is approximately the same
color as the Sab template of \cite{fuk95}.
The  $\tau<4.5 {\rm ~Gyr}$ selection criterion is $\simeq 0.15$ magnitudes
redder in $B_W-R$ than the $u^*-r^*>1.8$ (AB) color cut for SDSS red galaxies in 
\cite{zeh02}. A subsample, selected with $\tau<2.0 {\rm ~Gyr}$, was chosen to allow
estimates of clustering as a function of rest-frame color and 
comparisons with redder samples, including $z>0.8$ EROs. 
The rest-frame color of the $12~{\rm Gyr}$ old 
$\tau=2.0 {\rm ~Gyr}$  model with $E(B-V)=0.04$ intrinsic dust 
extinction is $B_W-R=1.77$, which is only $0.07$ magnitudes
bluer in $B_W-R$ than the $12~{\rm Gyr}$ old $\tau=0.6 {\rm ~Gyr}$ model.
The observed colors of the red galaxy sample, along with the 
$\tau$-models, are shown in Figure~\ref{fig:col}.
For comparison, the rest-frame colors of the \cite{col80} E and Sbc 
templates are $B_W-R=1.71$ and $B_W-R=1.14$. Objects with colors which 
differ significantly from the models can contaminate the sample. To reduce this 
contamination, only $0.5<R-I<1.6$ galaxies with photometric redshift fits 
with $\chi^2<3$ (including the prior) were included in the final sample.

\begin{figure}
\plottwo{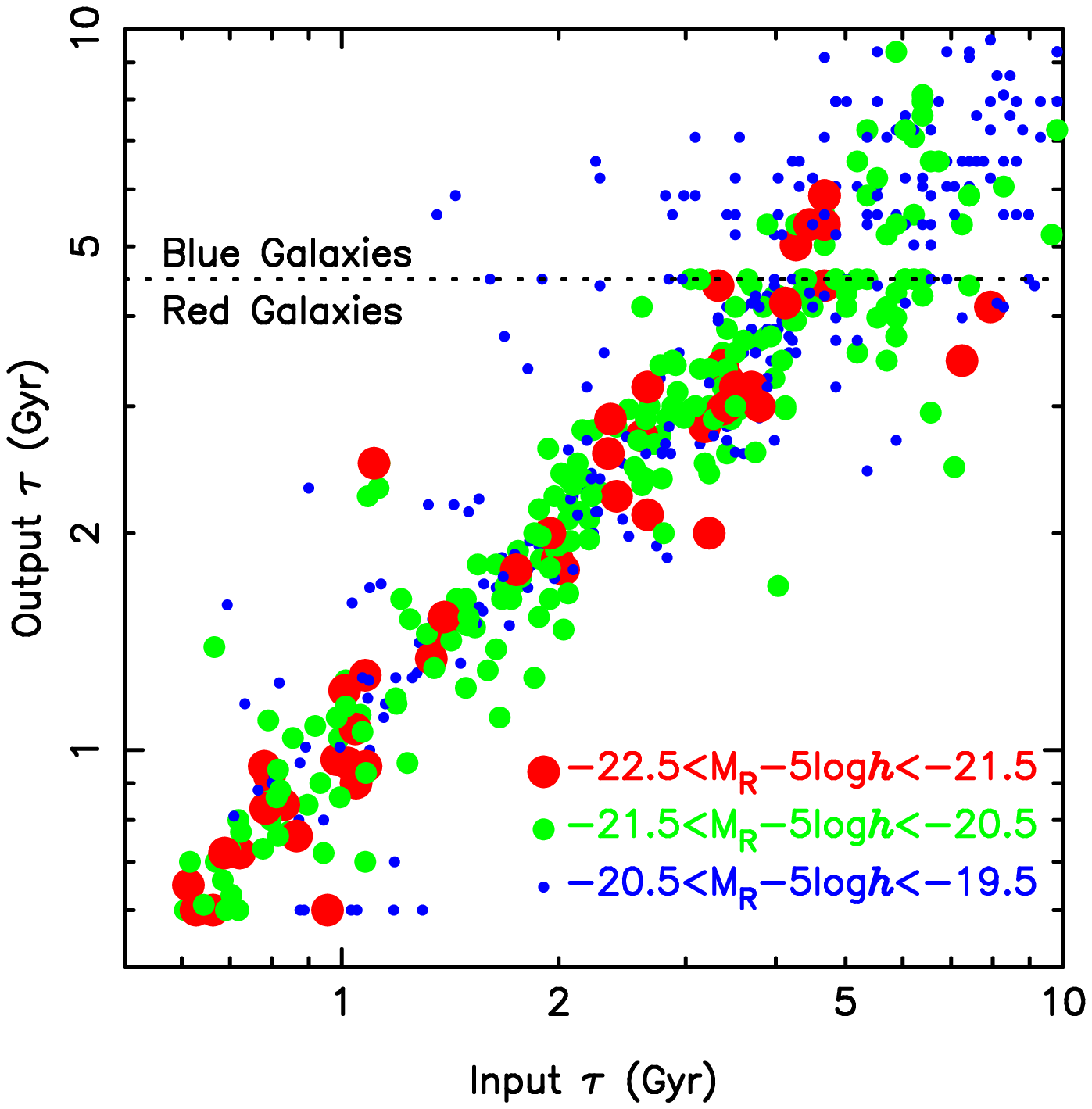}{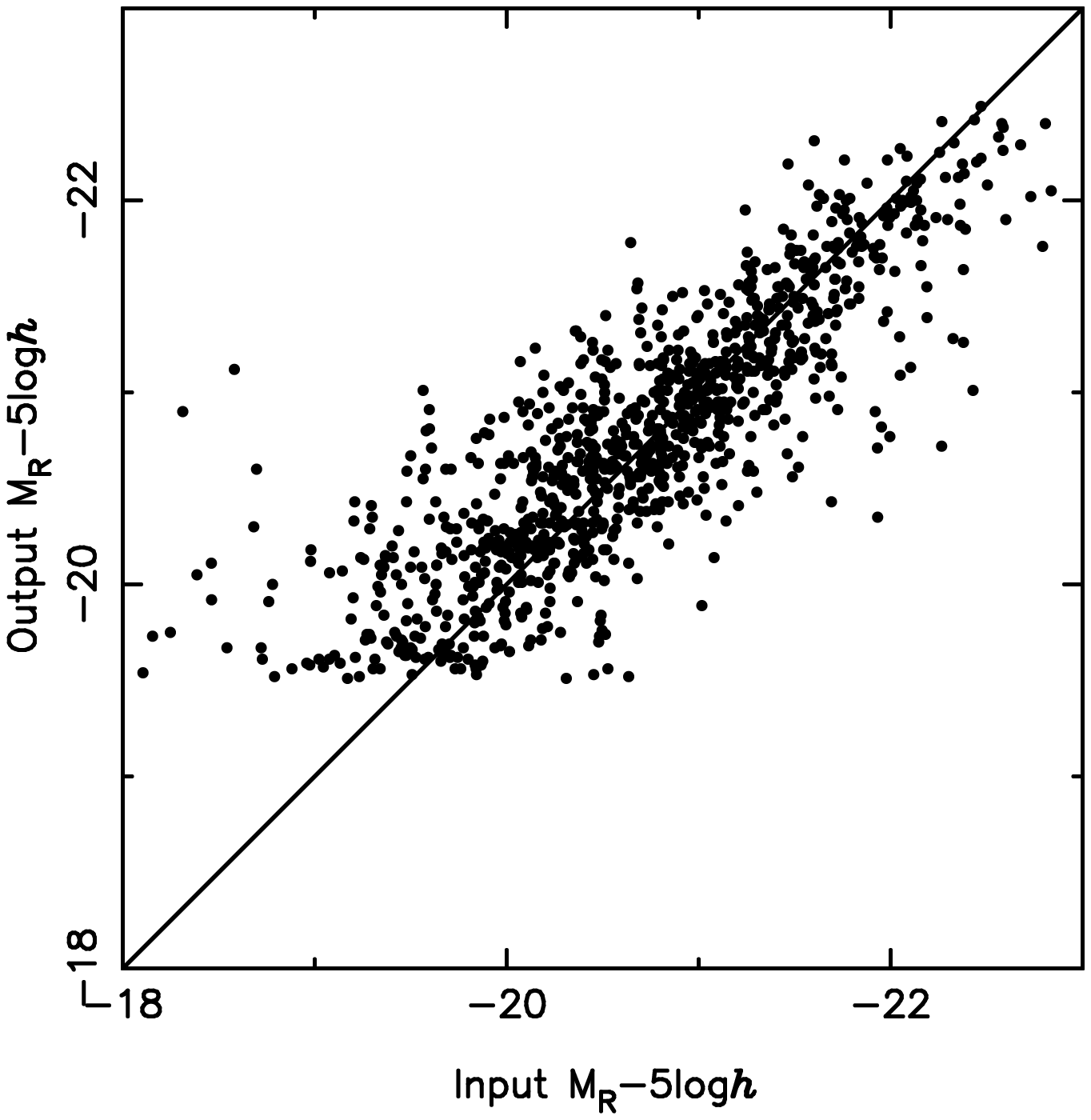}
\caption{A comparison of the input and measured spectral  types and absolute 
magnitudes for simulated galaxies with photometric redshifts in the range 
$0.30<z<0.90$. The $\tau<4.5~{\rm Gyr}$ selection criterion is shown with the 
dotted line in the left-hand panel. There is good agreement between the 
input and output $\tau$ and $M_R$ values for red galaxies. Few blue galaxies 
are scattered into the red galaxy sample. Some structure can be observed in the
$\tau$ output values due to the interpolated models only being evaluated 
when their colors and redshifts differ significantly from the other models.}
\label{fig:tau}
\end{figure}

The conclusions of this paper rely on the accuracy of the galaxy spectral 
classifications and absolute magnitudes. Significant contamination by late-type
spirals or $R>21$ faint blue galaxies will dramatically decrease angular and 
spatial two-point correlation functions \citep{efs91}. 
To minimize contamination, the photometric redshift range was restricted
to $0.30<z<0.90$, where the $\tau$-models are not degenerate for $B_WRI$ photometry.
The bulk of the spectra used in Figure~\ref{fig:photoz} have relatively low 
signal-to-noise ratios making it difficult to verify the selection criteria
spectroscopically with currently available datasets. Instead, we used the 
simulated data and compared the values of $\tau$ and $M_R$ used to generate
the model object with the output $\tau$ and $M_R$ values from the photometric redshift code.
As shown in Figure~\ref{fig:tau}, there is good agreement between the input and output 
values of $\tau$ and $M_R$ for red galaxies. Few blue galaxies are scattered into 
the red sample while a small fraction of red galaxies are scattered out 
of the sample. For the remainder of the paper, we restrict the magnitude
range to $R\leq 23.25$, where tests with simulated galaxies indicate
$\tau$ is being reliably measured. While some contamination is inevitable, 
it is reasonable to assume that weakly clustered faint blue galaxies \citep[e.g.,][]{efs91} 
are not dominating the measured angular correlation function. 
The final red galaxy sample contains 5325 objects, which is 14\% of all
39316, $R\leq23.25$ galaxies in the $1.2\Box^\circ$ sample area. 

\section{The correlation function}
\label{sec:cor}

We determined the angular correlation function using the \cite{lan93} estimator:
\begin{equation}
\hat\omega(\theta)=\frac{DD-2DR+RR}{RR}
\end{equation}
where $DD$, $DR$, and $RR$ are the number of galaxy-galaxy, galaxy-random 
and random-random pairs at angular separation $\theta\pm\delta\theta/2$.
The pair counts were determined in logarithmically spaced bins between 
$10^{\prime\prime}$ and $1^\circ$.

The random objects consist of copies of real galaxies which have had their 
positions changed to mimic objects that are randomly distributed across the 
sky. This does not result in a perfectly uniform surface density of objects across the 
field due to the completeness variations in the 4 subfields. When the ``random'' objects 
were distributed across the field, the probability of each object being detected in a 
given subfield was estimated and the product of this and the subfield area was used when 
determining which subfield the object would be placed. To decrease the contribution 
of the random objects to the shot-noise, 100 random object catalogs were generated 
and $DR$ and $RR$ were renormalized accordingly. 

The estimator of the correlation function is subject to the
integral constraint 
\begin{equation}
\int \int \hat \omega (\theta_{12}) d\Omega_1 d\Omega_2 \simeq 0
\end{equation}
\citep{gro77} which results in a systematic underestimate of the 
clustering. To remove this bias, the term
\begin{equation}
\omega_\Omega =
\frac{1}{\Omega^2} \int \int \omega (\theta_{12}) d\Omega_1 d\Omega_2
\end{equation}
was added to $\hat \omega (\theta)$ where $\Omega$ is the survey area.
The value of $\bar{n}^2\omega_\Omega$, where $\bar{n}$ is the mean number of galaxies per area
$\Omega$, is the contribution of clustering to the variance of the galaxy 
number counts \citep{gro77,efs91}. 
The angular correlation function was assumed to be a power-law given by 
\begin{equation}
\omega(\theta) = \omega(1^\prime) \left( \frac{\theta}{1^\prime} \right)^{1-\gamma} 
\end{equation}
where $\gamma$ is a constant. This is a good approximation of the 
observed spatial correlation function from the 2dFGRS and SDSS surveys
on scales of $\lesssim 10 h^{-1} {\rm Mpc}$ \citep{nor01,zeh02}. For a 
$\gamma=1.87$ power-law, the integral constraint for this study was approximately 
$6\%$ of the  amplitude of the correlation function at $1^\prime$.

We determined the covariance matrix of $\omega(\theta)$ using the approximation of 
\cite{eis01}:
\begin{equation}
C_\omega(\theta_i, \theta_j) = \frac{1}{\pi \Omega^2}
\int_0^\infty K P^2_2(K) J_0(K\theta_i) J_0(K\theta_j) dK
\label{eq:ez}
\end{equation}
where $J_0$ is a Bessel function and $P^2_2(K)$ is the
angular power spectrum,
\begin{equation}
P_2(K) = 2\pi \int_0^\infty  w(\theta) J_0(K\theta)\theta d\theta.
\end{equation} 
This approximation is best suited to correlation functions where 
$\omega(\theta)\ll 1$ and underestimates the covariance of very strongly
clustered objects. For the evaluation of $P_2(K)$, we truncated the 
power-law form of the correlation function at $2^\circ$ as the $z\sim 0$ galaxy correlation
function is $\approx 0$ on scales of $>20 h^{-1} {\rm Mpc}$ \citep[e.g.,][]{mad96, con02}.
However, the power-law fits to the data are only marginally affected by the value 
of $P_2(K)$ on large scales. If the angular correlation function bins have 
significant width, Equation~\ref{eq:ez} is modified to 
\begin{equation}
C_\omega(\theta_i, \theta_j) = 
\left(\frac{2}{\theta_{i,2}^2-\theta_{i,1}^2}\right)\left(\frac{2}{\theta_{j,2}^2-\theta_{j,1}^2}\right)
\int^{\theta_{i,2}}_{\theta_{i,1}}\int^{\theta_{j,2}}_{\theta_{j,1}}\theta \theta^\prime  
C_\omega(\theta,\theta^\prime) d\theta d\theta^\prime
\end{equation}
(D. Eisenstein 2003, private communication) where $\theta_1$ and $\theta_2$ are the inner and outer radii of the bins.
This can be rewritten as the single integral
\begin{eqnarray}
C_\omega(\theta_i, \theta_j) 
 = \frac{4}{(\theta_{i,2}^2-\theta_{i,1}^2)(\theta_{j,2}^2-\theta_{j,1}^2)\pi\Omega^2} \times \nonumber \\
 \int^\infty_0  P_2^2(K) \left[ \theta_{i,2}J_1(K\theta_{i,2})-\theta_{i,1}J_1
(K\theta_{i,1})\right] \left[\theta_{j,2}J_1(K\theta_{j,2})-\theta_{j,1}J_1(K\theta_{j,1})\right] \frac{dK}{K}.
\end{eqnarray}
The contribution of shot noise to the estimate of the covariance was included
by adding the reciprocal of the sky surface density of galaxies (per steradian) to $P_2(k)$. 
However, the shot noise only dominates the covariance on scales of $\lesssim 1^\prime$ 
for the red galaxy sample.

The spatial correlation function was obtained using Limber's (1954) equation:
\begin{equation}
\omega(\theta)= \int^\infty_0  \frac{dN}{dz}
\left[ \int^\infty_0 \xi (r(\theta,z,z^\prime),z) \frac{dN}{dz^\prime} dz^\prime \right] dz
\left/
\left( \int^\infty_0 \frac{dN}{dz} dz \right)^2 \right.
\end{equation}
where $\frac{dN}{dz}$ is the redshift distribution, $\xi$ is the spatial correlation function
and $r(\theta,z,z^\prime)$ is the comoving distance between two objects at redshifts $z$ 
and $z^\prime$ separated by angle $\theta$ on the sky. The spatial correlation function was assumed
to be a power law given by
\begin{equation}
\xi (r,z) = [ r/r_0(z)]^{-\gamma} 
\label{eq:spa}
\end{equation}
where 
\begin{equation}
r_0(z) = r_0(0) [1+z] ^{-(3+\epsilon-\gamma)/\gamma}
\label{eq:ep}
\end{equation}
and $\epsilon$ is a constant \citep{gro77}. Clustering is fixed in physical 
or comoving coordinates if $\epsilon=0$ or $\epsilon=\gamma-3$ respectively.

\begin{figure}
\plotone{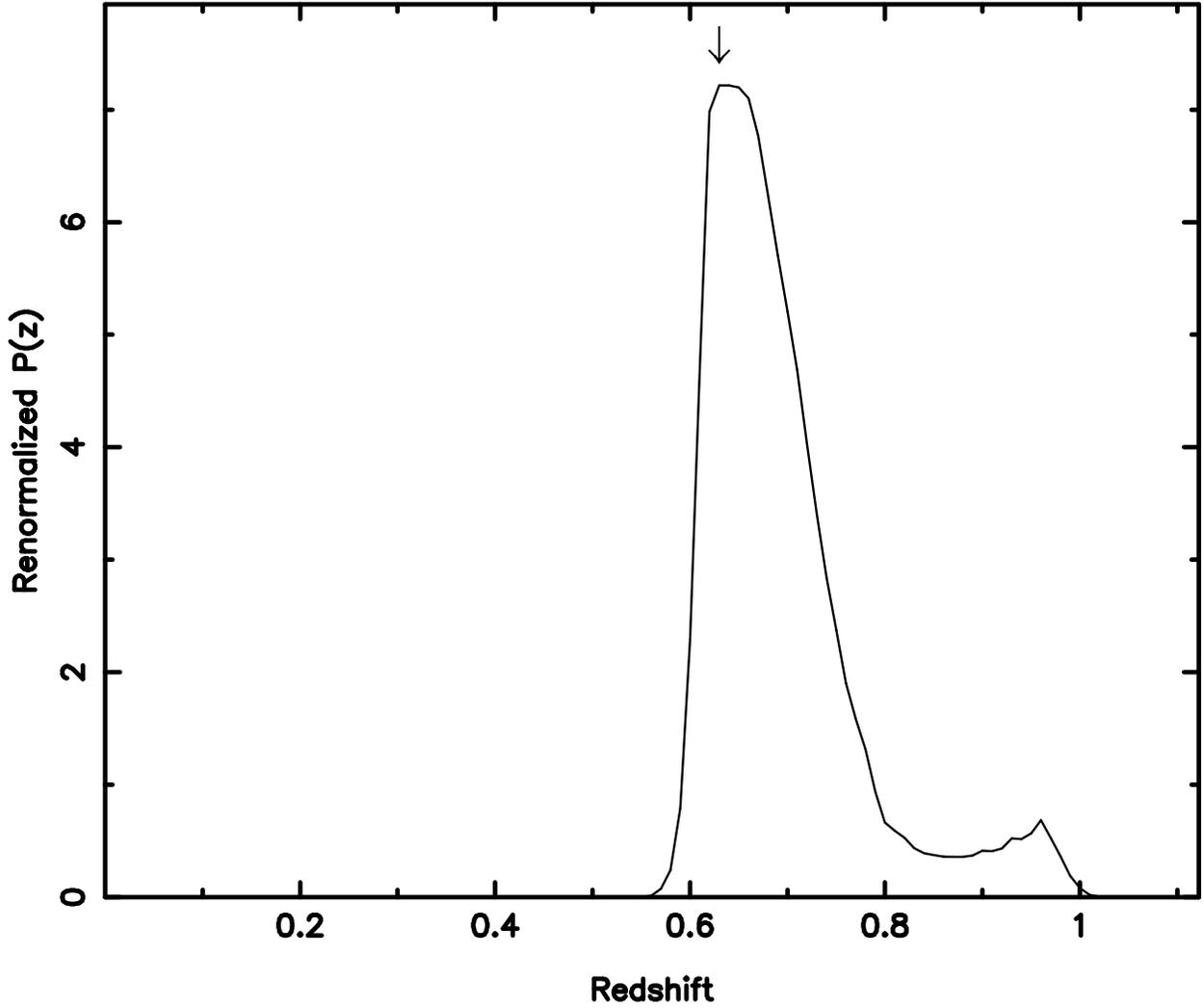}
\caption{The redshift likelihood distribution produced by the photometric redshift code.
${\rm NDWFS~J142518.9+353545}$, which has a spectroscopic redshift of 0.705, has a photometric redshift 
of 0.63 (shown by the arrow) and a spectral type of $\tau=0.64 ~{\rm Gyr}$. Unlike \cite{bru00} and 
\cite{bro01}, we can not adequately model the redshift likelihood distribution with a Gaussian.}
\label{fig:pz}
\end{figure}

\begin{figure}
\plottwo{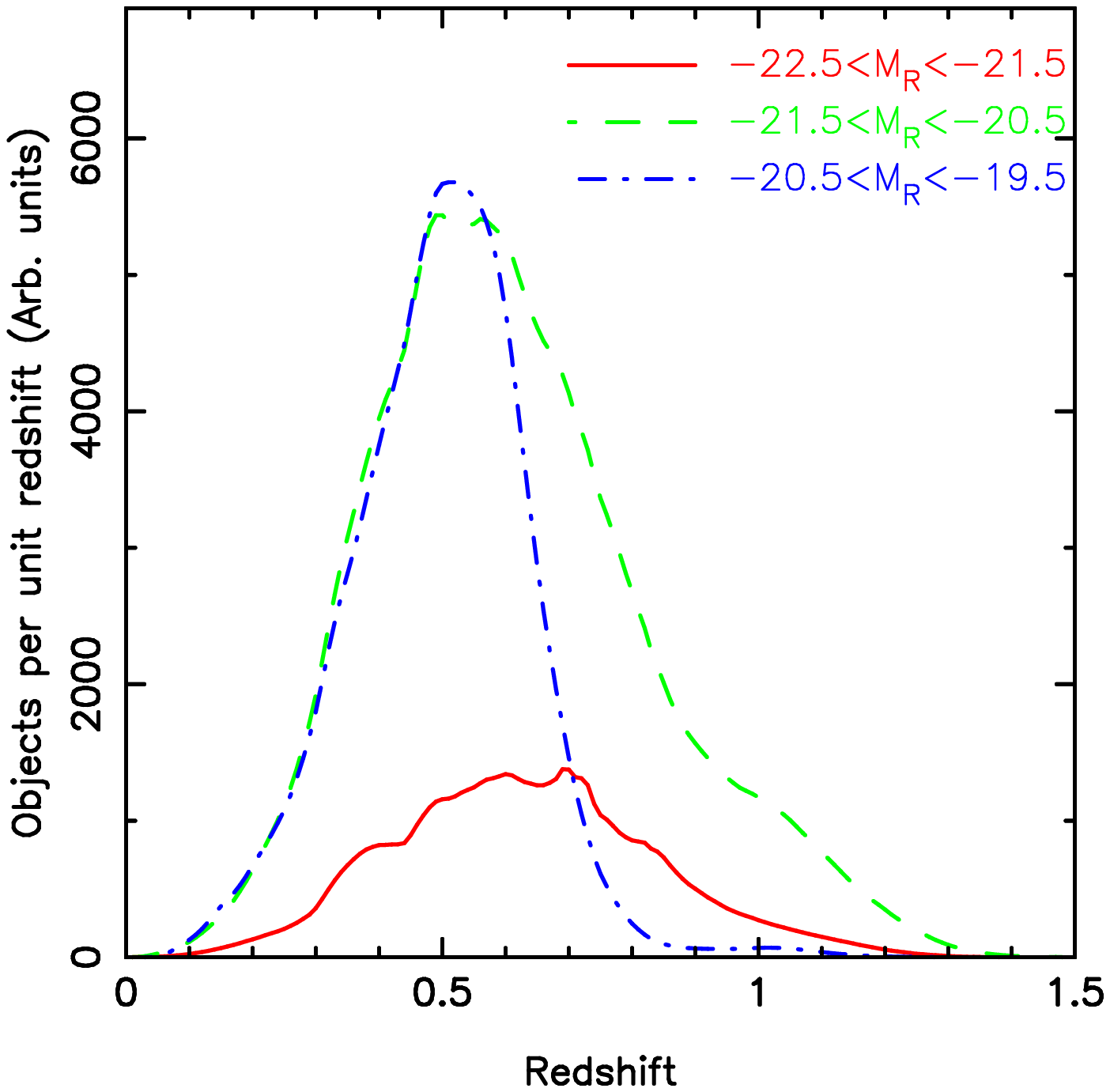}{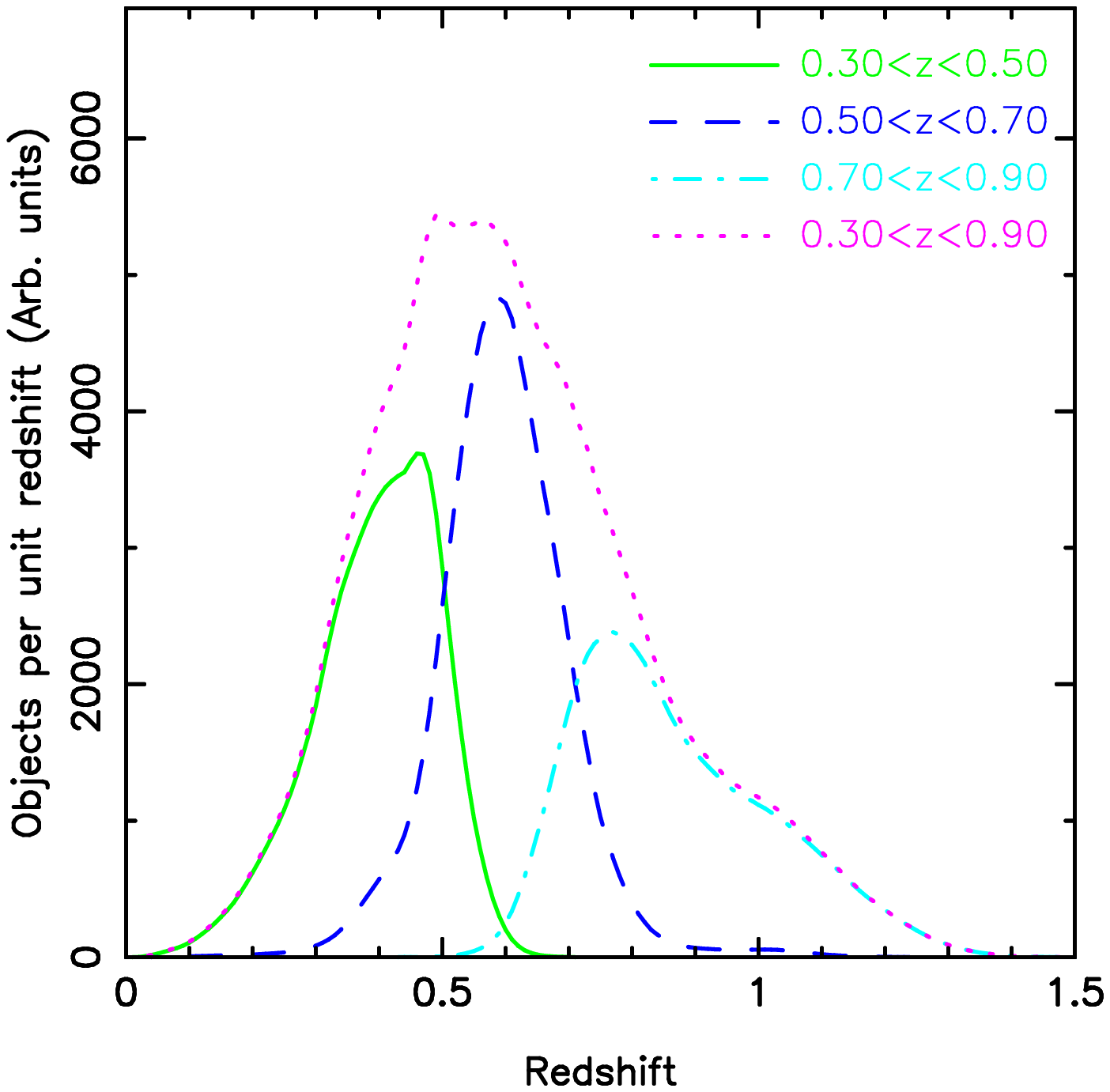}
\caption{The redshift distributions used to evaluate the spatial
correlation function of the $\tau<4.5 ~{\rm Gyr}$ samples as a function
of luminosity (left panel) and redshift (right panel). 
The redshift distributions of adjoining photometric redshift bins have significant 
overlaps and the resulting estimates of $r_0$ for these bins cannot be 
considered entirely independent. The redshift distribution of the most luminous red
galaxies shows evidence of individual structures in redshift space. 
The redshift distributions of the  $\tau<2.0 ~{\rm Gyr}$ subsamples 
(which are not shown) have comparable shapes but different normalizations. }
\label{fig:dndz}
\end{figure}

We estimated the redshift distribution for Limber's equation by 
summing the redshift likelihood distributions of the individual 
galaxies in each subsample. 
As shown in Figure~\ref{fig:pz}, the redshift distributions of individual 
galaxies can not be modeled with Gaussians and estimates derived from the 
photometric redshift code $\chi^2$ as a function of redshift must be used instead. 
Model redshift distributions for several subsamples selected
by luminosity and photometric redshift are shown in Figure~\ref{fig:dndz}. 

\section{The clustering of red galaxies}
\label{sec:clust}

\subsection{The angular and spatial correlation functions}

\begin{figure}
\plotone{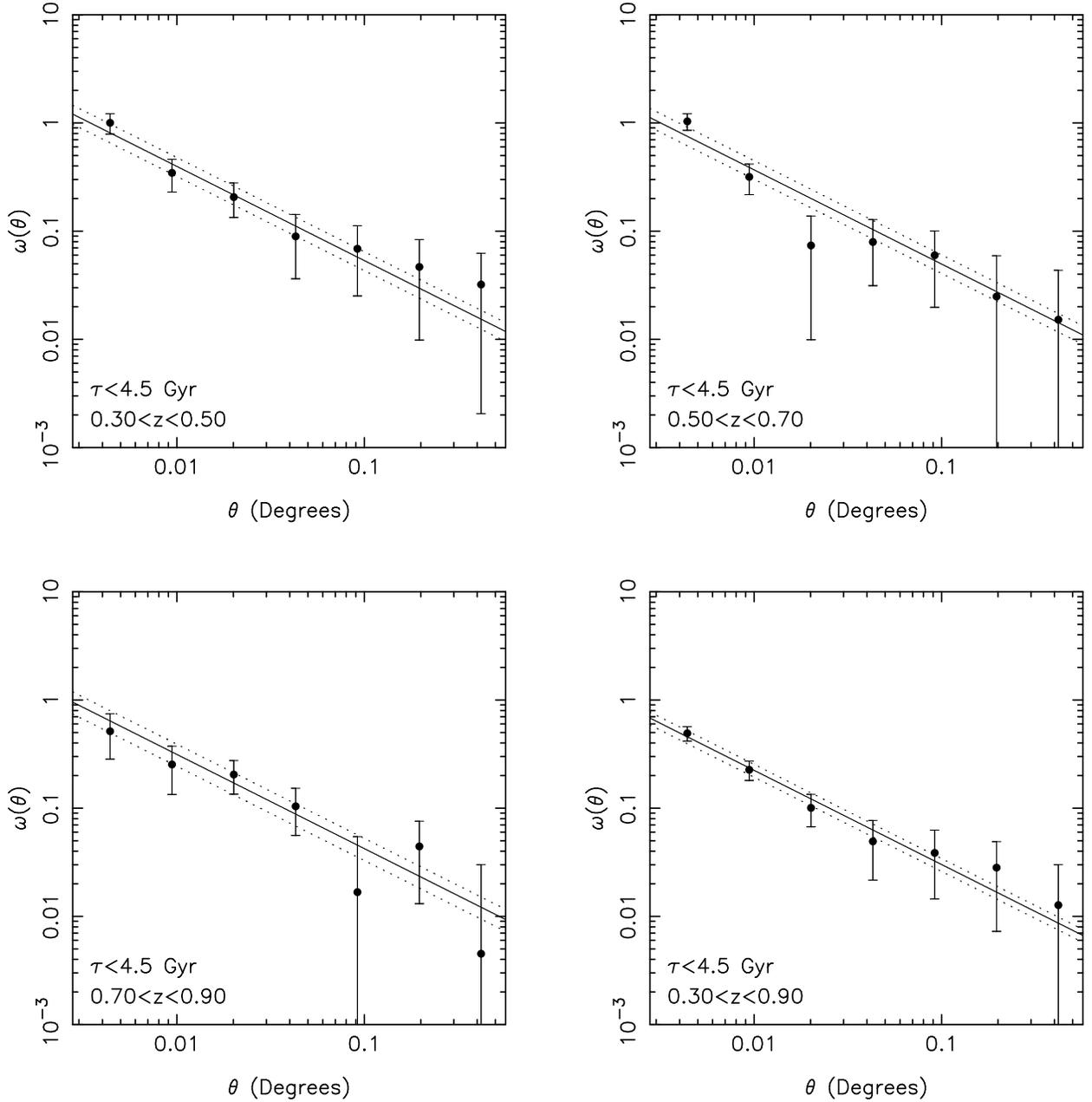}
\caption{The angular correlation functions of $\tau<4.5 ~{\rm Gyr}$, $-21.5<M_R-5{\rm log}h<-20.5$ 
galaxies selected from the NDWFS. The first three panels show the angular correlation 
functions for the narrowest photometric redshift bins while the last panel shows the 
angular correlation function for the $0.30<z<0.90$ bin. On small angular scales shot noise
dominates the uncertainties while on larger scales the contribution of clustering to the 
covariance dominates. Power-law fits to the data are shown along with $\pm 1 \sigma$ errors 
(dotted lines). For reference, $0.1^\circ$ corresponds to a transverse comoving 
distance of $3 h^{-1} {\rm Mpc}$ at $z\approx 0.7$.}
\label{fig:ang}
\end{figure}

The angular correlation function was determined for red galaxies in a series of
photometric redshift bins between $z=0.30$ and $z=0.90$, and absolute magnitude 
bins between $M_R-5{\rm log}h=-22.5$ and $M_R-5{\rm log}h=-19.5$. 
All bins are volume limited samples containing galaxies brighter than $R=23.25$.
A power-law of the form $\omega (\theta) = \omega(1^\prime) (\theta/1^\prime)^{1-\gamma}$ 
was fitted to the $\theta<0.25^\circ$ data with $\gamma$ fixed to $1.87$, the value 
for $z<0.15$ early-type galaxies \citep{nor02,zeh02}. We use $\omega(1^\prime)$ 
rather than $\omega(1^\circ)$ for the power-law fits as it depends less on the
assumed value of $\gamma$. The amplitude of the 
two-point angular correlation functions for these subsamples are
summarized in Tables~\ref{table:r0mag} and~\ref{table:r0z}. Angular 
correlation functions for $-21.5<M_R-5{\rm log}h<20.5$, $\tau<4.5~{\rm Gyr}$
galaxies are also plotted in Figure~\ref{fig:ang}.

A summary of the spatial clustering (parameterized by $r_0$) 
as a function of spectral type, absolute magnitude, and redshift is
presented in Tables~\ref{table:r0mag} and~\ref{table:r0z}. 
The estimates of $r_0$ in the narrow redshift bins are consistent with the values
in the widest redshift bins. While this is expected, it is not
always the case as the largest redshift bin contains object pairs which 
are not present in the smallest bin. The width of the redshift distribution
for the smallest bin strongly depends on the uncertainties of the photometric redshifts as
these uncertainties are comparable to the bin width. In contrast, the shape of the 
redshift distribution of the largest bin depends mostly on the bin width as the 
uncertainties of the photometric redshifts are smaller than the width of the bin. 
If the uncertainties are systematically underestimated, Limber's equation will
overestimate the number of close object pairs in the narrowest bin 
and underestimate the value of $r_0$. 
If the uncertainties in the redshift distribution were not included,
$r_0$ would vary between $3.9\pm 0.4 h^{-1} {\rm Mpc}$ and $5.5 \pm 0.4 h^{-1} {\rm Mpc}$
for the narrowest and widest redshift bins for $-21.0<M_R-5{\rm log}h<-20.0$, $\tau<4.5~{\rm Gyr}$ galaxies.
The measured evolution of $r_0$ should be independent of the width of the redshift 
bins and confirming this is an extremely useful internal consistency check 
which should always be applied to correlation functions using photometric redshifts.

\subsection{Clustering as a function of absolute magnitude}
\label{sec:abs}

\begin{figure}
\plotone{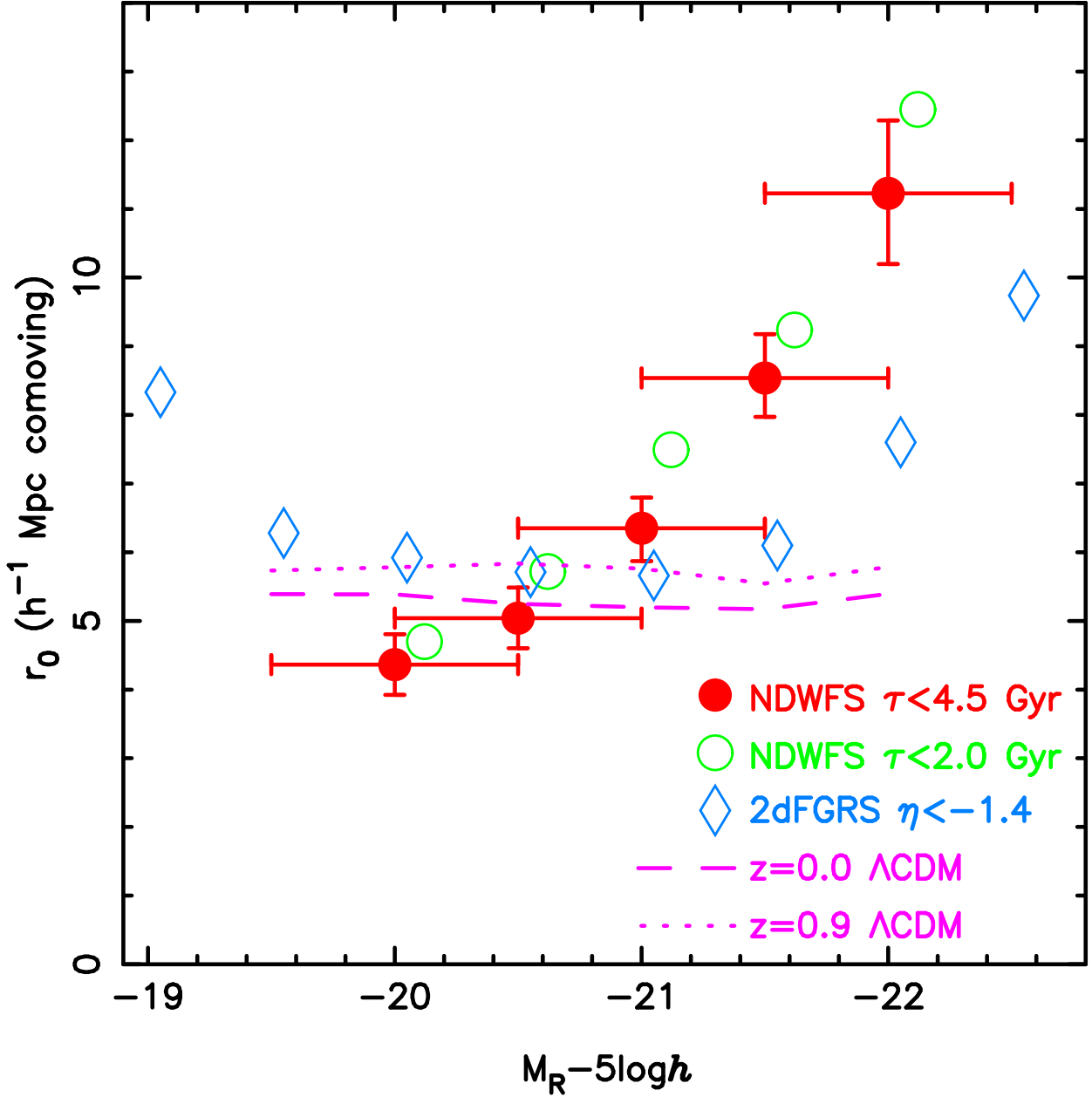}
\caption{The spatial clustering of red galaxies as a function of absolute magnitude.
For clarity, only the uncertainties for the $\tau < 4.5~{\rm Gyr}$ data are shown and the
$\tau<2.0~{\rm Gyr}$ data has been shifted slightly to the right. While datapoints
brighter than $M_R-5{\rm log}h=-20.5$ include galaxies over the entire $0.30<z<0.90$
photometric redshift range, fainter bins have a truncated redshift range.
The $z<0.15$ 2dFGRS data points, which have uncertainties of $\sim 0.5 h^{-1} {\rm Mpc}$, 
are over-plotted using the assumption that $B_J-R=1.25$ for red galaxies. The 
clustering of red galaxies is a strong function of luminosity and a weaker function 
of spectral type. A $\Lambda$CDM simulation of the clustering $B-R>1.24$ galaxies 
\citep[][A. Benson 2002, private communication]{ben01} reproduces the clustering of 
$L^*$ galaxies but the $141^3 h^{-3} {\rm Mpc^3}$ simulation volume is 
insufficient to measure the clustering of luminous galaxies.}
\label{fig:r0mag}
\end{figure}

Figure~\ref{fig:r0mag} and Table~\ref{table:r0mag} present 
estimates of the spatial clustering of red galaxies as a function 
of evolution corrected absolute magnitude. Absolute magnitude bins
containing galaxies brighter than $M_R-5{\rm log}h=-20.5$ include the
entire $0.30<z<0.90$ photometric redshift range while fainter bins
have truncated redshift ranges which are listed in Table~\ref{table:r0mag}.
While redder galaxies are more strongly clustered than bluer galaxies, the 
striking correlation is between $r_0$ and absolute magnitude.
While there is a mild correlation with luminosity at 
$M_R-5{\rm log}h\sim -20.0$, the value of $r_0$ rapidly increases from 
$r_0=6.3 h^{-1} {\rm Mpc}$ at $M_R-5{\rm log}h=-21$ to 
$r_0=11.2 h^{-1} {\rm Mpc}$ at $M_R-5{\rm log}h=-22$. 
Similar behavior is seen for at $z<0.15$ in the SSRS2 and 2dFGRS
\citep{wil98,nor02}, and, with lower significance, in CNOC2 at 
$z<0.4$ \citep{she01}. \cite{hog03} observe similar trends in the 
SDSS at $0.05<z<0.22$ by measuring density of galaxy neighbors 
within $8 h^{-1} {\rm Mpc}$ spheres as a function of galaxy 
color and luminosity. \cite{wil03} measures 
$r_0=4.02\pm 0.22 h^{-1} {\rm Mpc}$ for $0.2<z<1.0$, $M_B<-18.61$ 
red galaxies, but her model redshift distribution
does not include the uncertainties of her $V-I$ photometric redshifts,
so her value of $r_0$ is a lower limit. A summary of previous measurements
of red galaxy correlation functions is provided in 
Table~\ref{table:prev}.
 
\begin{figure}
\plotone{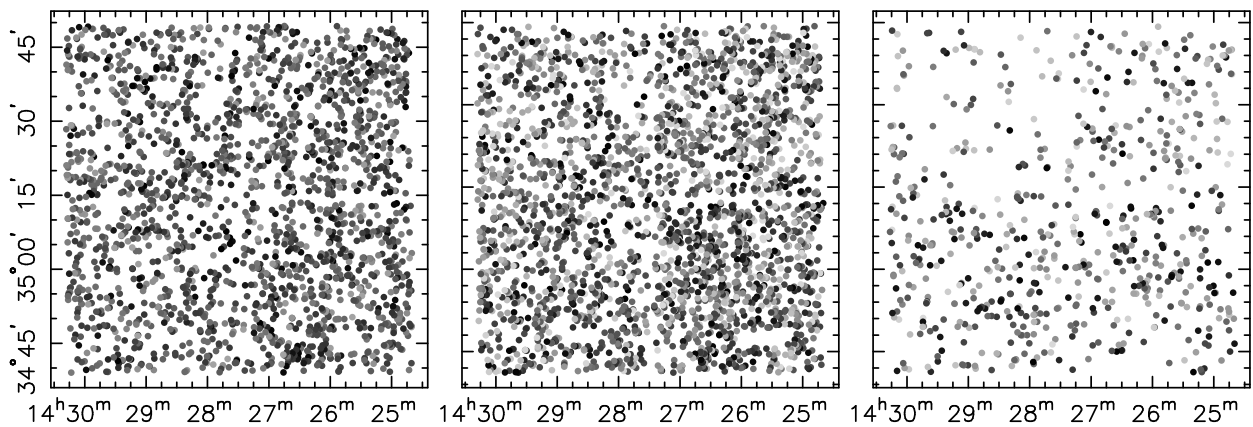}
\caption{The distribution of red galaxies on the plane of the sky for several absolute
magnitude bins. The bins are (from left to right) 
$-20.5<M_R-5{\rm log}h <-19.5$, $-21.5<M_R-5{\rm log}h <-20.5$ and $-22.5<M_R-5{\rm log}h <-21.5$.
The greyscale of the dots is related to the photometric redshift, 
with dark dots at low redshift and light dots at high redshift. Rectangular voids
in the galaxy distribution are the boundaries between the NDWFS subfields and regions
surrounding bright stars. The distribution of luminous galaxies clearly 
shows evidence for structures comparable to the area of the subfields, suggesting that
surveys of this size are susceptible to cosmic variance.} 
\label{fig:dist}
\end{figure}

While the $r_0$ values of $M_R-5{\rm log} h>-21.5$ galaxies from the 
NDWFS and 2dFGRS show only marginal differences, the $r_0$ values of luminous 
galaxies in the NDWFS are higher than those of the 2dFGRS with $\approx 2\sigma$ 
significance. Clustering evolution would be expected to produce decreasing $r_0$ 
values with increasing redshift rather than the opposite trend seen in 
Figure~\ref{fig:r0mag}. However, as luminous galaxies are more strongly 
clustered than $L^*$ galaxies, estimates of their clustering are also more susceptible 
to cosmic variance. The distribution of galaxies on the plane of the sky, 
which is plotted in Figure~\ref{fig:dist}, clearly shows that luminous galaxies are in 
structures comparable in size to the field-of-view. While it is plausible that 
selection effects could produce the observed structures, stars selected with the
same selection criteria do not show similar large-scale structure. 
In addition, in Figure~\ref{fig:dndz}, the model redshift distribution of 
the most luminous red galaxies shows evidence of individual structures.
We therefore assume that the $2\sigma$ difference between 
the clustering of luminous galaxies in the NDWFS and 2dFGRS is due 
cosmic variance. Even if cosmic variance were not an issue, it is 
difficult to measure clustering evolution with galaxies in the 
absolute magnitude range where $r_0$ is strongly correlated with luminosity 
as small luminosity errors can translate into large errors in $r_0$. 
It is therefore preferable to measure clustering evolution with 
$M_R-5{\rm log}h>-21.5$ galaxies, as we have done in \S\ref{sec:z}.

As the correlation between $r_0$ and spectral type is relatively 
weak in the NDWFS and 2dFGRS, our measurements of the evolution of 
clustering are not sensitive to small errors in the estimates of the 
spectral types. The strong correlation between $r_0$ and absolute magnitude for luminous galaxies is a
prediction of recent large volume $\Lambda$CDM simulations \citep[e.g.,][]{ben01}. 
As shown in Figure~\ref{fig:r0mag}, a $141^3 h^{-3} {\rm Mpc^3}$ $\Lambda$CDM 
simulation with galaxy selection criteria similar to the $\tau<4.5 ~{\rm Gyr}$ 
sample \citep[][A. Benson 2002, private communication]{ben01} is a good approximation of the 
clustering of $M_R\sim -21$ red galaxies.

\subsection{Evolution of the spatial correlation function}
\label{sec:z}

The evolution of spatial clustering galaxies was studied with $-21.5<M_R-5{\rm log}h<-20.5$ 
red galaxies. Luminous galaxies were excluded as their spatial clustering is 
strongly correlated with absolute magnitude and small redshift or spectral
evolution errors could produce large changes in the measured spatial clustering.
The faint limit was chosen to allow the same range of absolute magnitudes 
to be studied over a broad redshift range.

\begin{figure}
\plotone{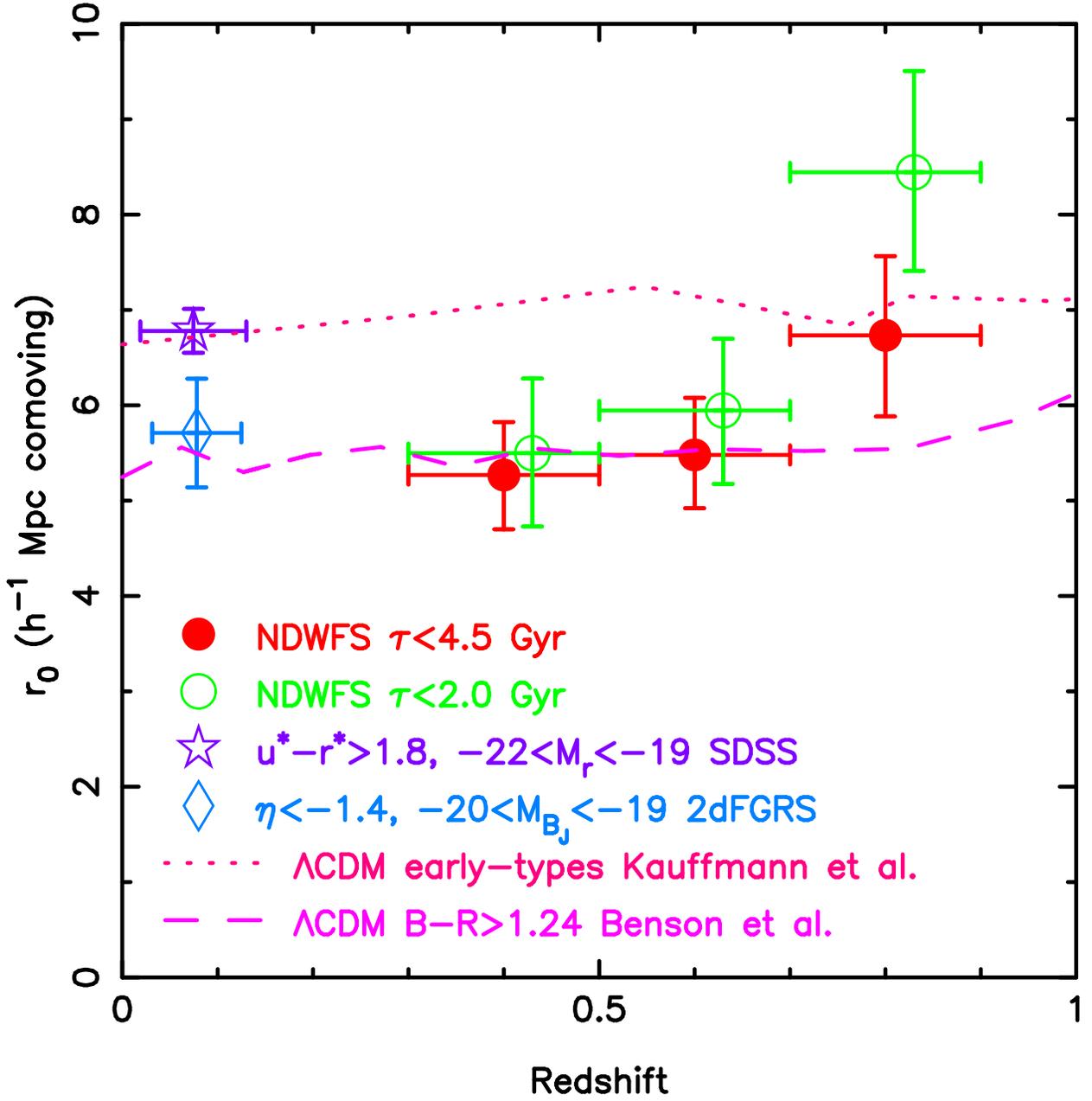}
\caption{The spatial correlation function parameter $r_0$ as a function of redshift
for the NDWFS, 2dFGRS, and SDSS red galaxy samples. For clarity, 
the $\tau< 2.0~{\rm Gyr}$ data points have been shifted slightly to the right. 
The horizontal bars indicate the photometric (NDWFS) and spectroscopic (2dFGRS, SDSS) 
redshift ranges. The clustering of $\tau<4.5~{\rm Gyr}$ galaxies 
(which were selected to allow direct comparison with the 2dFGRS)
is well approximated by $\Lambda$CDM simulations of the 
clustering of early-type and $B-R>1.24$ galaxies \citep[A. Benson 2002, private communication]{kau99,ben01}. 
From the plot, it is clear the value of $r_0$ (in comoving coordinates) undergoes
little or no evolution over the redshift range observed.}
\label{fig:r0}
\end{figure}

As shown in Figure~\ref{fig:r0} and Table~\ref{table:r0z}, no significant evolution 
of $r_0$ (comoving) occurs over the redshift range studied. Two models
of the clustering of red galaxies, derived from the GIF $\Lambda$CDM 
simulations \citep{jen98}, are plotted in Figure~\ref{fig:r0} and 
provide good approximations to the measured clustering from the 
NDWFS, SDSS and 2dFGRS. The \cite{kau99} model predicts the clustering 
of early-type galaxies with stellar masses of $>3\times 10^{10}M_{\sun}$ while 
the Benson (2002, private communication) simulation models the clustering of 
$B-R>1.24$, $-20.5<M_R-5{\rm log}h<-21.5$ galaxies.
If the evolution of the underlying dark matter distribution at 
$z<1$ is well described by the linear or quasi-linear growth of 
density perturbations, the bias of red galaxies must be rapidly 
evolving with redshift.

\begin{figure}
\plottwo{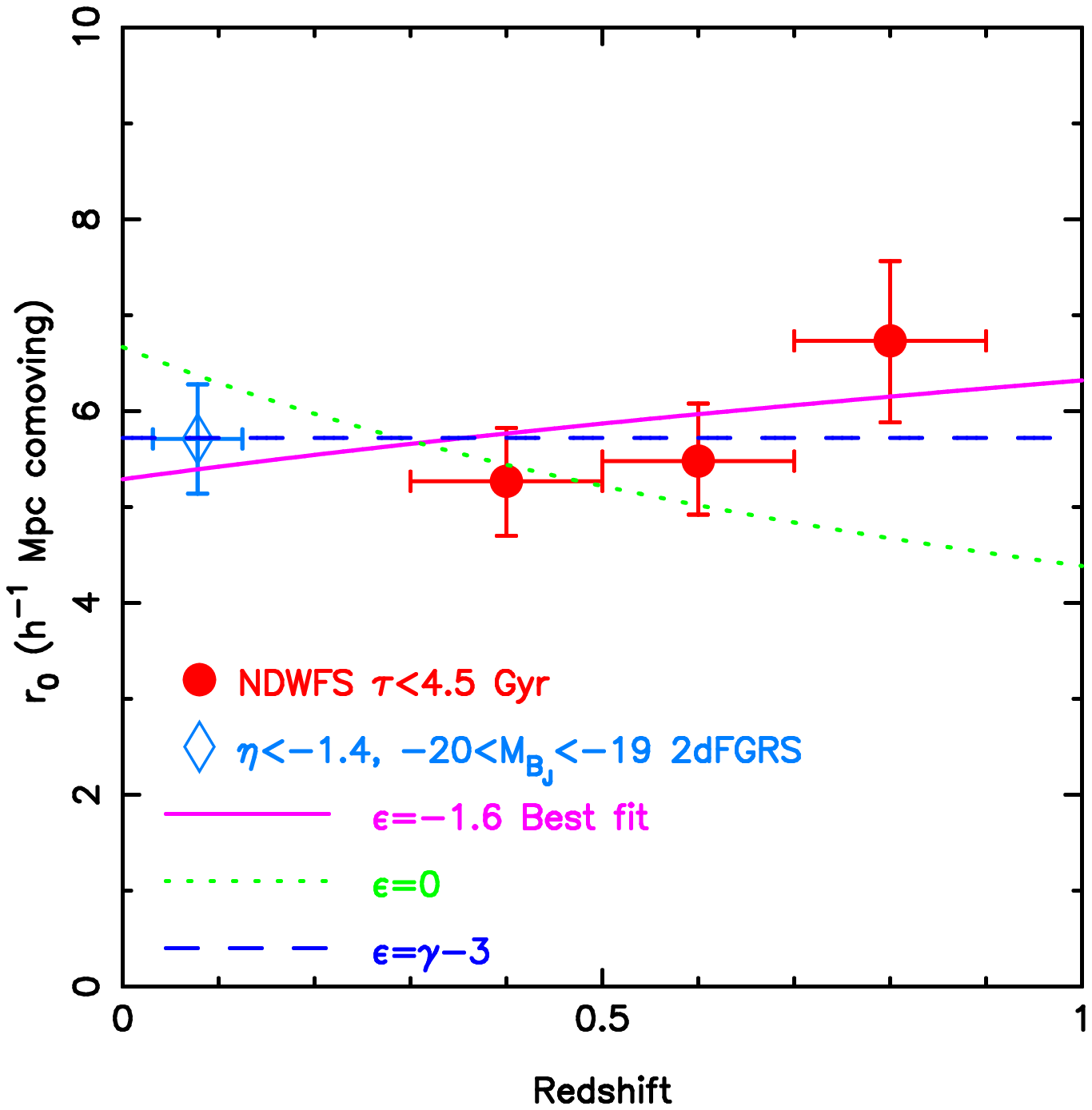}{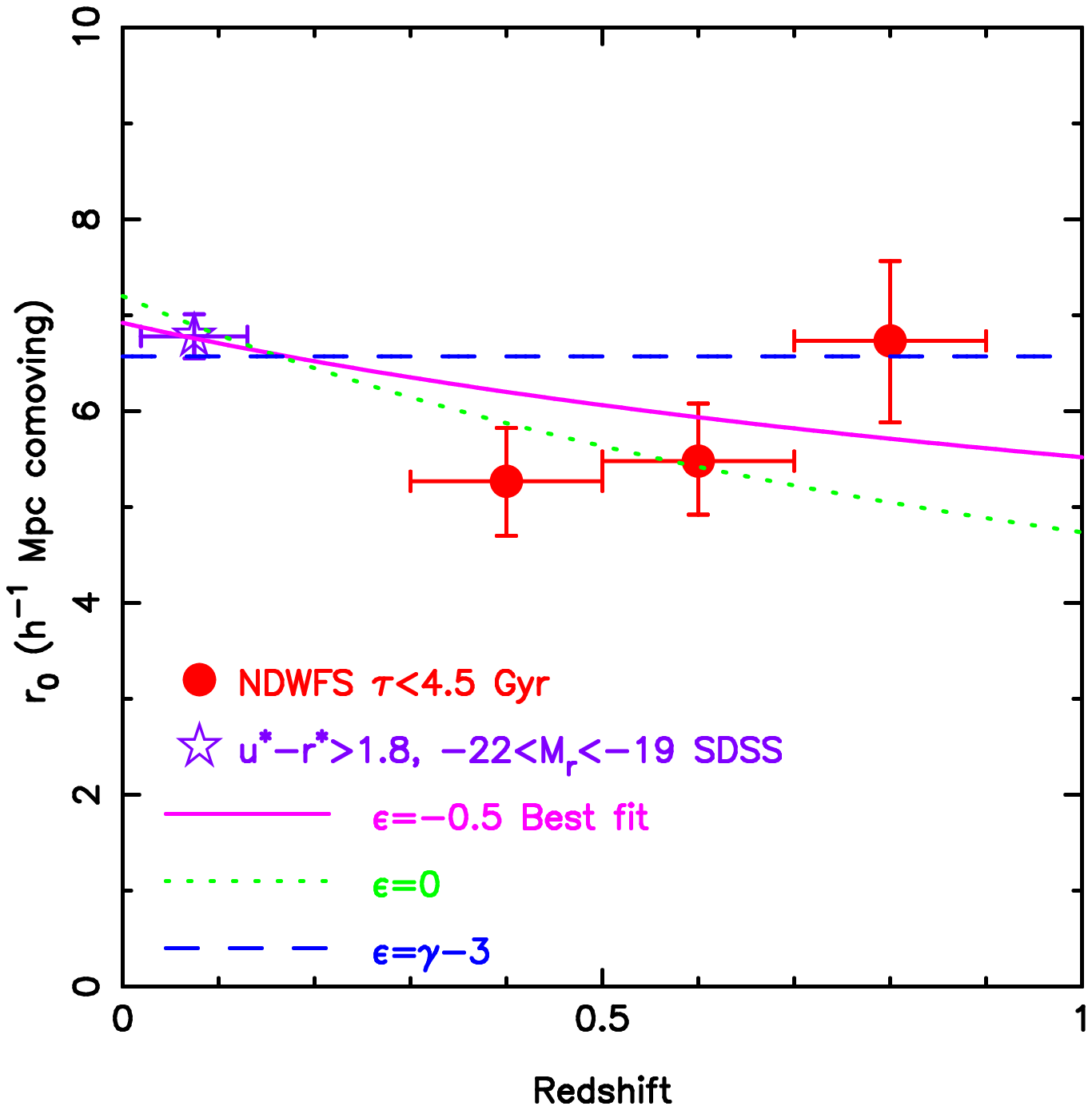}
\caption{The evolution of $r_0$ (comoving). The NDWFS data are shown with the comparable
low redshift data from the 2dFGRS (left panel) and SDSS (right panel). 
Models with clustering fixed in physical coordinates ($\epsilon=0$) are rejected with $2.5\sigma$ 
confidence by the NDWFS and 2dFGRS while models with little or no evolution ($\epsilon\simeq\gamma-3$) 
provide acceptable fits to the observed clustering.} 
\label{fig:epev}
\end{figure}

The evolution of $r_0$ was empirically measured by estimating the 
clustering evolution parameter $\epsilon$ (from Equation~\ref{eq:ep}). 
The $0.30<z<0.50$ and $0.70<z<0.90$ bins were assumed to be independent 
estimates of $r_0$ at the median redshifts of their model redshift 
distributions. The 2dFGRS and SDSS clustering estimates were also 
included to provide additional constraints on $\epsilon$. 
The selection criteria for the NDWFS $\tau<4.5~{\rm Gyr}$ sample
allow direct comparison with the 2dFGRS early-type sample of
\cite{nor02}. The present-day colors of the NDWFS $\tau<4.5~{\rm Gyr}$ sample
are only $\simeq0.15$ magnitudes redder in $B_W-R$ than the 
SDSS $u^*-r^*>1.8$ sample of \cite{zeh02}.
The best-fit estimates of $\epsilon$ are summarized in Table~\ref{table:ep} 
and plotted in Figure~\ref{fig:epev}. Models with $\epsilon=0$ are rejected with $>2.5\sigma$ 
confidence when constraints from the 2dFGRS are included with the NDWFS data.
The evolution of $r_0$ is consistent with clustering fixed in 
comoving coordinates ($\epsilon=\gamma-3$) at $z<0.90$.

\section{Discussion}
\label{sec:dis}

The clustering of red galaxies undergoes little or no evolution from 
$z\sim 0.9$ to the present-day. At $z>0.9$, the only 
published clustering study using comparable template selection 
criteria is that of \cite{fir02} which used galaxies fitted with the 
evolving E and Sbc templates of the HYPERZ photometric redshift code 
\citep{bol00}. Their estimate of $r_0 = 7.0 \pm 1.9 h^{-1}{\rm Mpc}$, 
with $\gamma$ assumed to be $1.8$, is consistent with $r_0$ remaining fixed in comoving 
coordinates to $z\approx 1.5$. However, to differentiate between currently 
plausible $\Lambda$CDM models larger survey areas, on the order of the complete
NDWFS ($\approx 18 \Box^\circ$), will be required. 

Further constraints on the clustering of $1<z<2$ galaxies are 
available from samples of QSOs \citep{cro01} and $R-K>5$ extremely red 
objects \citep[EROs;][]{dad01,fir02,roc02}. The QSO spatial correlation function, 
measured in redshift space ($s_0$ replacing $r_0$), marginally increases from 
$s_0=5.28\pm^{0.72}_{0.89} h^{-1} {\rm Mpc}$  with $\gamma=1.72\pm^{0.23}_{0.22}$ 
at $z<0.95$ to $s_0=6.93\pm^{1.32}_{1.64} h^{-1} {\rm Mpc}$ with 
$\gamma=1.64\pm^{0.29}_{0.27}$ at  $2.10<z<2.90$ \citep{cro01}. However, as 
pointed out by \cite{cro01}, interpretation of QSO clustering relies on 
poorly constrained models of QSOs lifetimes and host populations as a 
function of redshift.

EROs are easier to relate to low redshift populations than QSOs as the majority 
appear to be the progenitors of early-type galaxies with the remainder
being dusty starbursts \citep{dey99,lid00,mor00}. EROs could therefore 
extend the redshift range of early-type spatial correlation function 
evolution estimates to $z\approx 1.5$. With the exception of 
$H<20.5$ EROs in the Las Campanas Infrared Survey \citep{fir02}, 
all EROs clustering studies measure $r_0\geq10 h^{-1} {\rm Mpc}$ 
\citep{dad01,fir02,roc02}. This is comparable to the clustering of 
the most luminous red and early-type galaxies at $z<1$.

If $K>19$ EROs correspond to $\lesssim L^*$ galaxies \citep{dad01,roc02}, 
the ERO clustering results are difficult to reconcile with $z<1$ 
studies and $\Lambda$CDM theory. The clustering of red $L^*$ galaxies in 
the NDWFS can be increased to $r_0\sim 9 h^{-1} {\rm Mpc}$ only by increasing 
the photometric redshift $1\sigma$ uncertainties to $\sim 40\%$ of the 
photometric redshifts or increasing the stellar contamination to $\sim 40\%$. 
These scenarios are inconsistent with the redshift comparisons in 
Figure~\ref{fig:photoz} and the lack of a stellar locus
in Figure~\ref{fig:col}. The discrepancy between the ERO $r_0$ values 
and $z<1$ studies could be due to errors in the ERO model 
redshift distributions used to deproject the angular correlation function. 
\cite{fir02}, who measure $r_0=7.7\pm 2.4 h^{-1} {\rm Mpc}$ for $H<20.5$ EROs, have few 
$z\gtrsim 1.5$ objects  while the \cite{dad01} and 
\cite{roc02} models have a significant fraction of EROs at $z\approx 2$.
Improved constraints on ERO clustering should be provided with model 
redshift distributions constrained with photometric redshifts which 
have been verified with spectroscopic samples.
The clustering of EROs, as measured from the NDWFS, will be described 
in a future paper (Brown et al., in preparation) using a $B_WRIK$ dataset.

\section{Summary}
\label{sec:sum}

We have used the NOAO Deep Wide-Field Survey to measure 
the clustering of $0.30 \lesssim z \lesssim 0.90$ red galaxies.
The wide-field and $B_WRI$ bands allow large 
galaxy samples to be selected as a function of spectral type 
and absolute magnitude using photometric redshifts.
PEGASE2 spectral evolution models with exponentially decreasing star formation rates 
have been used to select an evolutionary sequence of related galaxies as a function of 
redshift. The red sample,  with present-day rest-frame colors of $B_W-R>1.44$,
was chosen to allow direct comparison with the 
low redshift early-type sample from the 2dFGRS.
The clustering of red galaxies is strongly correlated with luminosity, 
with $r_0$ increasing from $4.4\pm 0.4 h^{-1} {\rm Mpc}$ at 
$M_R-5{\rm log}h\approx -20.0$ to $11.2\pm 1.0 h^{-1} {\rm Mpc}$ 
at $M_R-5{\rm log}h\approx -22.0$. 
Clustering evolution measurements with samples where the
distribution of spectral types and luminosities are a function
of redshift will be dominated by selection effects.
The strength of $r_0$ (comoving) as a function of absolute magnitude in our sample 
is comparable to estimates at $z<0.15$ from the 2dFGRS, with differences at 
high luminosity being attributable to structures of sizes comparable to the
field-of-view. No significant evolution of $r_0$ was detected in comparisons of the 
NDWFS with the 2dFGRS and SDSS. For $0.30<z<0.90$, $-21.5<M_R-5{\rm log}h<-20.5$, 
$B_W-R>1.44$ galaxies, the largest sample studied, the value of $r_0$ is 
$6.3\pm 0.5 h^{-1} {\rm Mpc}$ with $\gamma$ fixed at $1.87$. 
The strong clustering and lack of detectable evolution appears consistent with 
recent $\Lambda$CDM models where the bias undergoes rapid evolution and
$r_0$ undergoes little evolution at $z<2$.

\acknowledgments

This research was supported by the National Optical Astronomy Observatory which is 
operated by the Association of Universities for Research in Astronomy (AURA), Inc. 
under a cooperative agreement with the National Science Foundation.
We thank our colleagues on the NDWFS team and the KPNO and CTIO observing support staff.
We thank  Frank Valdes, Lindsey Davis and the IRAF team for the MSCRED and astrometry 
packages used to reduce the mosaic imaging data. 
S. Croom, S. Dawson, R. Green, S. Malhotra, J. Rhoads, P. Smith, H. Spinrad, D. Stern and S. Warren kindly 
provided spectroscopic redshifts for galaxies in the Lockman Hole and NDWFS prior to publication.
We thank Daniel Eisenstein
for productive discussions concerning the uncertainties of correlation functions.
Andrew Benson kindly provided estimates of the clustering
of red galaxies derived from the GIF $\Lambda$CDM simulation.
Valerie Mikles' research was supported by the NOAO/KPNO REU Program, 
funded by the National Science Foundation. 
This research has made use of the NASA/IPAC Extragalactic Database, 
which is operated by the Jet Propulsion Laboratory, California Institute 
of Technology, under contract with the National Aeronautics and Space Administration.

\clearpage

\clearpage

\begin{deluxetable}{cccccccccccc}
\tabletypesize{\scriptsize}
\tablecaption{The four subfields of the first NDWFS public data release.
\label{table:fields}}
\tablehead{
\colhead{Subfield name} &
\colhead{R.A.}  &
\colhead{Decl.} &
\multicolumn{3}{c}{Delivered FWHM ($^{\prime\prime}$)\tablenotemark{a}} & 
\multicolumn{3}{c}{Integration time(hours)} &
\multicolumn{3}{c}{50\% completeness limit\tablenotemark{b}} \\
&
\colhead{(J2000.0)} &
\colhead{(J2000.0)} &
\colhead{$B_W$} & \colhead{$R$} & \colhead{$I$} &
\colhead{$B_W$} & \colhead{$R$} & \colhead{$I$} &
\colhead{$B_W$} & \colhead{$R$} & \colhead{$I$} 
}
\startdata
NDWFS J1426+3531 & 14~26~00.8 & +35~31~32 & 1.2 & 1.6 & 1.3 & 2.1 & 1.7 & 3.0 & 26.8 & 24.8 & 24.7 \\
NDWFS J1426+3456 & 14~26~01.4 & +34~56~32 & 1.3 & 1.2 & 1.0 & 2.3 & 1.2 & 2.1 & 26.5 & 25.3 & 24.9 \\
NDWFS J1428+3531 & 14~28~52.8 & +35~31~39 & 1.5 & 1.3 & 0.7 & 2.3 & 1.7 & 2.2 & 26.4 & 25.0 & 25.6 \\
NDWFS J1428+3456 & 14~28~52.2 & +34~56~39 & 1.6 & 1.3 & 1.3 & 2.3 & 1.7 & 2.6 & 26.2 & 25.0 & 23.6 \\
\enddata
\tablenotetext{a}{Measured using the SExtractor FWHM values of bright unsaturated stars.}
\tablenotetext{b}{Determined with SExtractor using artificial stellar objects inserted into copies of the real data.}
\end{deluxetable}

\begin{deluxetable}{ccccccccc}
\rotate
\tabletypesize{\scriptsize}
\tablecaption{The correlation function of red galaxies in the NDWFS as a function of absolute magnitude.
\label{table:r0mag}}
\tablewidth{0pt}
\tablehead{
\colhead{Selection} & 
\colhead{Photo. $z$ range} & 
\colhead{Absolute magnitude} &
\colhead{Apparent magnitude} &
\colhead{Galaxies} & 
\colhead{Contamination} & 
\colhead{$\omega(1^\prime)$} & 
\colhead{Median $z$} & 
\colhead{$r_0 (h^{-1} {\rm Mpc})$}
}
\startdata
$\tau<4.5 ~{\rm Gyr}$ & 0.30-0.90 & $-22.50<M_R<-21.50$ & $18.18\leq R\leq 22.43$ &  660 & 0.009 & $0.41 \pm 0.07$ & 0.64 & $11.2 \pm 1.0$ \\
$\tau<4.5 ~{\rm Gyr}$ & 0.30-0.90 & $-22.00<M_R<-21.00$ & $18.30\leq R\leq 23.01$ & 1677 & 0.005 & $0.24 \pm 0.03$ & 0.61 & $8.5 \pm 0.6$ \\
$\tau<4.5 ~{\rm Gyr}$ & 0.30-0.90 & $-21.50<M_R<-20.50$ & $18.76\leq R\leq 23.25$ & 2651 & 0.023 & $0.14 \pm 0.02$ & 0.60 & $6.3 \pm 0.5$ \\
$\tau<4.5 ~{\rm Gyr}$ & 0.30-0.75 & $-21.00<M_R<-20.00$ & $19.17\leq R\leq 23.17$ & 2429 & 0.009 & $0.12 \pm 0.02$ & 0.55 & $5.0 \pm 0.4$ \\
$\tau<4.5 ~{\rm Gyr}$ & 0.30-0.65 & $-20.50<M_R<-19.50$ & $19.72\leq R\leq 23.08$ & 1756 & 0.025 & $0.12 \pm 0.02$ & 0.51 & $4.4 \pm 0.4$ \\
\\
$\tau<2.0 ~{\rm Gyr}$ & 0.30-0.90 & $-22.50<M_R<-21.50$ & $18.18\leq R\leq 22.43$ &  488 & 0.011 & $0.52 \pm 0.10$ & 0.66 & $12.4 \pm 1.2$ \\
$\tau<2.0 ~{\rm Gyr}$ & 0.30-0.90 & $-22.00<M_R<-21.00$ & $18.30\leq R\leq 23.01$ & 1122 & 0.006 & $0.29 \pm 0.05$ & 0.64 & $9.2 \pm 0.8$ \\
$\tau<2.0 ~{\rm Gyr}$ & 0.30-0.90 & $-21.50<M_R<-20.50$ & $18.76\leq R\leq 23.25$ & 1592 & 0.037 & $0.20 \pm 0.03$ & 0.62 & $7.5 \pm 0.6$ \\
$\tau<2.0 ~{\rm Gyr}$ & 0.30-0.75 & $-21.00<M_R<-20.00$ & $19.25\leq R\leq 23.17$ & 1337 & 0.016 & $0.16 \pm 0.04$ & 0.57 & $5.7 \pm 0.7$ \\
$\tau<2.0 ~{\rm Gyr}$ & 0.30-0.65 & $-20.50<M_R<-19.50$ & $19.87\leq R\leq 23.08$ &  929 & 0.043 & $0.14 \pm 0.04$ & 0.53 & $4.7 \pm 0.7$ \\
\enddata
\end{deluxetable}


\begin{deluxetable}{ccccccccc}
\rotate
\tabletypesize{\scriptsize}
\tablecaption{The correlation function of red galaxies in the NDWFS as a function of photometric redshift.
\label{table:r0z}}
\tablewidth{0pt}
\tablehead{
\colhead{Selection} & 
\colhead{Photo. $z$ range} & 
\colhead{Absolute magnitude} &
\colhead{Apparent magnitude} &
\colhead{Galaxies} & 
\colhead{Contamination} & 
\colhead{$\omega(1^\prime)$} & 
\colhead{Median $z$} & 
\colhead{$r_0 (h^{-1} {\rm Mpc})$}
}
\startdata
$\tau<4.5 ~{\rm Gyr}$ & 0.30-0.50 & $-21.50<M_R<-20.50$ & $18.76\leq R\leq 21.25$ &  853 & 0.003 & $0.25 \pm 0.05$ & 0.42 & $5.3 \pm 0.6$ \\
$\tau<4.5 ~{\rm Gyr}$ & 0.50-0.70 & $-21.50<M_R<-20.50$ & $20.05\leq R\leq 22.50$ & 1023 & 0.002 & $0.24 \pm 0.05$ & 0.60 & $5.5 \pm 0.6$ \\
$\tau<4.5 ~{\rm Gyr}$ & 0.70-0.90 & $-21.50<M_R<-20.50$ & $20.99\leq R\leq 23.25$ &  775 & 0.071 & $0.20 \pm 0.05$ & 0.85 & $6.7 \pm 0.8$ \\
$\tau<4.5 ~{\rm Gyr}$ & 0.30-0.70 & $-21.50<M_R<-20.50$ & $18.76\leq R\leq 22.50$ & 1876 & 0.003 & $0.19 \pm 0.03$ & 0.52 & $5.9 \pm 0.5$ \\
$\tau<4.5 ~{\rm Gyr}$ & 0.50-0.90 & $-21.50<M_R<-20.50$ & $20.05\leq R\leq 23.25$ & 1798 & 0.032 & $0.17 \pm 0.03$ & 0.69 & $6.3 \pm 0.5$ \\
$\tau<4.5 ~{\rm Gyr}$ & 0.30-0.90 & $-21.50<M_R<-20.50$ & $18.76\leq R\leq 23.25$ & 2651 & 0.023 & $0.14 \pm 0.02$ & 0.60 & $6.3 \pm 0.5$ \\
\\
$\tau<2.0 ~{\rm Gyr}$ & 0.30-0.50 & $-21.50<M_R<-20.50$ & $18.76\leq R\leq 21.25$ &  435 & 0.005 & $0.31 \pm 0.08$ & 0.44 & $5.5 \pm 0.8$ \\
$\tau<2.0 ~{\rm Gyr}$ & 0.50-0.70 & $-21.50<M_R<-20.50$ & $20.12\leq R\leq 22.50$ &  665 & 0.004 & $0.29 \pm 0.07$ & 0.61 & $5.9 \pm 0.8$ \\
$\tau<2.0 ~{\rm Gyr}$ & 0.70-0.90 & $-21.50<M_R<-20.50$ & $21.20\leq R\leq 23.25$ &  492 & 0.111 & $0.31 \pm 0.07$ & 0.85 & $8.4 \pm 1.0$ \\
$\tau<2.0 ~{\rm Gyr}$ & 0.30-0.70 & $-21.50<M_R<-20.50$ & $18.76\leq R\leq 22.50$ & 1100 & 0.004 & $0.20 \pm 0.04$ & 0.55 & $5.8 \pm 0.7$ \\
$\tau<2.0 ~{\rm Gyr}$ & 0.50-0.90 & $-21.50<M_R<-20.50$ & $20.12\leq R\leq 23.25$ & 1157 & 0.050 & $0.22 \pm 0.04$ & 0.69 & $7.1 \pm 0.7$ \\
$\tau<2.0 ~{\rm Gyr}$ & 0.30-0.90 & $-21.50<M_R<-20.50$ & $18.76\leq R\leq 23.25$ & 1592 & 0.037 & $0.20 \pm 0.03$ & 0.62 & $7.5 \pm 0.6$ \\
\enddata
\end{deluxetable}

\begin{deluxetable}{ccccccc}
\rotate
\tabletypesize{\scriptsize}
\tablecaption{A summary of several studies of early-type and red galaxy correlation functions.
\label{table:prev}}
\tablewidth{0pt}
\tablehead{
\colhead{Survey\tablenotemark{a,b}} & 
\colhead{Redshift Range} &  
\colhead{Galaxies} & 
\colhead{Magnitude Range} & 
\colhead{Selection} & 
\colhead{$r_0 (h^{-1}{\rm  Mpc~comoving})$\tablenotemark{c,d}} & 
\colhead{$\gamma$\tablenotemark{e}}
}
\startdata
NDWFS    & $0.30<z<0.90      $ & 2651  & $-21.5<M_R<-20.5$ & $\tau<4.5~{\rm Gyr}$ & $6.3\pm 0.5$ & $1.87$ \\  
NDWFS    & $0.30<z<0.90      $ & 1592  & $-21.5<M_R<-20.5$ & $\tau<2.0~{\rm Gyr}$ & $7.5\pm 0.6$ & $1.87$ \\  
\\
Perseus-Pisces & $z \leq 0.04          $ & 278    & $M_{Z_W}<-19.5$    & Morphology 	     & $8.35\pm 0.75$ & $2.05^{+0.10}_{-0.08}$ \\
SSRS2  & $z \leq 0.020                 $ & 395    & $M_B<-19.4$        & Morphology 	     & $5.10\pm 0.38$ & $1.91\pm 0.26$ \\
SSRS2  & $z \leq 0.027                 $ & 418    & $M_B<-20.0$        & Morphology 	     & $5.27\pm 0.46$ & $1.86\pm 0.29$ \\
SSRS2  & $z \leq 0.033                 $ & 372    & $M_B<-20.5$        & Morphology 	     & $5.73\pm 0.56$ & $2.30\pm 0.46$ \\
SSRS2  & $z \leq 0.040                 $ & 272    & $M_B<-20.9$        & Morphology 	     & $8.60\pm 1.44$ & $2.45\pm 0.71$ \\
APM    & $z \lesssim 0.1               $ & 336    & $-22<M_{B_J}<-15 $ & Morphology   	     & $5.9\pm 0.7$   & $1.85\pm 0.13$ \\
2dFGRS & $0.016 \leq z \leq 0.071      $ & 1909   & $-18.5<M_{B_J}<-17.5 $ & Red SED ($\eta<-1.4$) & $8.33\pm 1.82$ & $1.87\pm 0.23$ \\ 
2dFGRS & $0.020 \leq z \leq 0.086      $ & 3717   & $-19.0<M_{B_J}<-18.0 $ & Red SED ($\eta<-1.4$) & $6.28\pm 1.46$ & $1.98\pm 0.11$ \\
2dFGRS & $0.031 \leq z \leq 0.125      $ & 10,135 & $-20.0<M_{B_J}<-19.0 $ & Red SED ($\eta<-1.4$) & $5.71\pm 0.57$ & $1.87\pm 0.09$ \\
2dFGRS & $0.048 \leq z \leq 0.150      $ & 6434   & $-21.0<M_{B_J}<-20.0 $ & Red SED ($\eta<-1.4$) & $6.10\pm 0.72$ & $1.80\pm 0.12$ \\
2dFGRS & $0.072 \leq z \leq 0.150      $ & 686    & $-22.0<M_{B_J}<-21.0 $ & Red SED ($\eta<-1.4$) & $9.74\pm 1.16$ & $1.95\pm 0.37$ \\
SDSS   & $0.019\leq z\leq  0.13        $ & 19,603 & $14.5<r^*< 17.6$ & Rest-frame $u^*-r^*>1.8$ & $6.78\pm 0.23$ & $1.86\pm 0.03$ \\ 
\\
PDF    & $z \lesssim 0.5               $ & 22,359 & $B_J <22.5$ & Redder than CWW Sbc	& $6.0\pm 0.3$   & 1.9 \\
K20    & $0.5 \lesssim z \lesssim  2.0 $ & 400   & $K   <19.2$ & $R-K>5$             	& $12\pm  3$     & 1.8 \\
CNOC2  & $0.120 < z < 0.270 $ & 248 & $R_C<21.5$, $M_R<-20$ & Red CWW templates & $5.35\pm 0.20$ & $2.05\pm 0.08$ \\
CNOC2  & $0.270 < z < 0.382 $ & 234 & $R_C<21.5$, $M_R<-20$ & Red CWW templates & $6.55\pm 1.16$ & $2.05\pm 0.08$ \\
CNOC2  & $0.382 < z < 0.510 $ & 238 & $R_C<21.5$, $M_R<-20$ & Red CWW templates & $6.99\pm 0.57$ & $2.10\pm 0.11$ \\
CNOC2  & $0.12 < z < 0.40 $ & 254   & $-20.00<M_R<-19.25$ & Red CWW templates  & $5.82\pm 0.81$ & $1.85\pm 0.08$ \\
CNOC2  & $0.12 < z < 0.40 $ & 276   & $-20.65<M_R<-20.00$ & Red CWW templates  & $5.40\pm 0.19$ & $1.89\pm 0.06$ \\
CNOC2  & $0.12 < z < 0.40 $ & 278   & $-22.52<M_R<-20.65$ & Red CWW templates  & $6.71\pm 0.53$ & $2.14\pm 0.02$ \\
UH8K   & $0.20<z<0.90$      & 3382  & $M_B\lesssim-18.61$ & Redder than CWW Sbc & $4.02\pm 0.22$ & 1.8 \\ 
LCIRS  & $0.3 \lesssim z \lesssim  0.8 $ & 272   & $H   <20.5$ & Evolving E \& Sbc 	& $7.0\pm 1.6$   & 1.8 \\
LCIRS  & $0.8 \lesssim z \lesssim  1.5 $ & 355   & $H   <20.5$ & Evolving E \& Sbc 	& $7.0\pm 1.9$   & 1.8 \\
LCIRS  & $0.7 \lesssim z \lesssim  1.5 $ & 337   & $H   <20.0$ & $R-H>4$             	& $11.1\pm 2.0$  & 1.8 \\
LCIRS  & $0.7 \lesssim z \lesssim  1.5 $ & 312   & $H   <20.5$ & $R-H>4$             	& $7.7\pm 2.4$   & 1.8 \\
ELAIS N2 & $0.5 \lesssim z \lesssim  2.0 $ & 166   & $K   <21.0$      & $R-K>5$             & 10-13          & 1.8 \\
\enddata
\tablenotetext{a}{Perseus-Pisces \citep{guz97}, SSRS2 \citep{wil98}, APM \citep{lov95}, 2dFGRS \citep{nor02}, SDSS \citep{zeh02},
PDF \citep{bro01}, K20 \citep{dad01}, CNOC2 \citep{she01}, UH8K \citep{wil03}, LCIRS \citep{fir02}, and ELAIS N2 \citep{roc02}.}
\tablenotetext{b}{For clarity the surveys have been split into 3 groups; NDWFS, $z<0.2$, and $z>0.2$.}
\tablenotetext{c}{For other studies, uncertainties are as published and may not include the effect of the covariance on the uncertainty estimates.}
\tablenotetext{d}{The parameter $r_0^\gamma/1.73$ is used instead of $r_0$ for CNOC2 \citep{she01}.}
\tablenotetext{e}{Where the value of $\gamma$ was fixed, the value is given without an error estimate.}
\end{deluxetable}

\begin{deluxetable}{ccccc}
\tabletypesize{\scriptsize}
\tablecaption{Constraints on $\epsilon$ (as defined in Equation \ref{eq:ep}) provided by the NDWFS, SDSS, and 2dFGRS.\label{table:ep}}
\tablewidth{0pt}
\tablehead{
\colhead{NDWFS Sample} & \colhead{$z<0.2$ Sample\tablenotemark{a}} & \colhead{$r_0(0) (h^{-1}{\rm  Mpc~comoving})$} & \colhead{$\epsilon$}
}
\startdata
$\tau<2.0 ~{\rm Gyr}$ & none                      & $3.1\pm_{1.0}^{1.3}$\tablenotemark{b} & $-4.2\pm_{1.3}^{1.3}$ \\ 
$\tau<4.5 ~{\rm Gyr}$ & none                      & $3.8\pm_{1.0}^{1.3}$\tablenotemark{b} & $-2.9\pm_{1.1}^{1.2}$ \\ 
$\tau<4.5 ~{\rm Gyr}$ & SDSS $u^*-r^*>1.8$, $-22.0<M_{r^*}<-19.0$  & $6.9\pm_{0.3}^{0.3}$ & $-0.5\pm_{0.4}^{0.5}$ \\ 
$\tau<4.5 ~{\rm Gyr}$ & 2dFGRS $\eta<-1.4$, $-20.0<M_{B_J}<-19.0$  & $5.3\pm_{0.6}^{0.7}$ & $-1.6\pm_{0.6}^{0.6}$ \\ 
\enddata
\tablenotetext{a}{The median redshifts of the SDSS and 2dFGRS samples are assumed to be 0.1. We use the
uncertainties of \cite{nor02} and \cite{zeh02} as published.}
\tablenotetext{b}{As the NDWFS red galaxy sample photometric redshift range is $0.30<z<0.90$,
the value of $r_0(0)$ from the NDWFS alone is an extrapolation, 
which is strongly dependent on the best-fit value of $\epsilon$.}
\end{deluxetable}

\end{document}